\documentclass[aps,prb,twocolumn,showpacs,superscriptaddress,10pt,longbibliography]{revtex4-2}

\usepackage{mathtools}
\usepackage[colorlinks=true,linkcolor=red]{hyperref}
\usepackage[sort&compress]{natbib}
\usepackage{graphicx}
\usepackage{physics}
\usepackage{xcolor}
\usepackage[normalem]{ulem}
\usepackage{orcidlink}

\newcommand{\bohumadd}{Theoretische Physik III, Ruhr-Universität Bochum, Bochum 44801, Germany}
\newcommand{\pcsadd}{Center for Theoretical Physics of Complex Systems, Institute for Basic Science(IBS), Daejeon, Korea, 34126}
\newcommand{\ustadd}{Basic Science Program, Korea University of Science and Technology (UST), Daejeon 34113, Republic of Korea}

\newcommand{\ec}{E_\mathrm{c}}
\newcommand{\ej}{E_\mathrm{J}}
\newcommand{\fcr}{f_\mathrm{c}}

\newcommand{\heff}{H_\mathrm{eff}}
\newcommand{\ueff}{U_\mathrm{eff}}
\newcommand{\keff}{K_\mathrm{eff}}

\newcommand{\kbt}{\hbar/(k_\mathrm{B}T)}
\newcommand{\onf}{\omega_\mathrm{nf}}
\newcommand{\ecz}{E_\mathrm{c0}}
\newcommand{\gff}{G_\mathrm{f}}

\newcommand{\vx}{\vec{r}}
\newcommand{\vy}{\vec{r}^\prime}

\newcommand{\alexei}[1]{\textcolor{brown}{#1}}

\newcommand{\BG}{\mathcal{G}}

\newcommand{\BC}{\mathcal{C}}

\newcommand{\BZ}{\mathcal{Z}}

\newcommand{\kcr}{K_\mathrm{cr}}

\begin{document}

\title{Collective quantum phases in frustrated arrays of Josephson junctions}

\author{M. V. Fistul\,\orcidlink{0000-0002-0265-2534
}}
    \affiliation{\bohumadd}

\author{O. Neyenhuys\,\orcidlink{0000-0002-8715-5181}}
    \affiliation{\bohumadd}

\author{B. Pernack\,\orcidlink{0009-0007-7682-1197}}
    \affiliation{\bohumadd}

\author{I. M. Eremin\,\orcidlink{0000-0003-0557-8015}}
    \affiliation{\bohumadd}

\author{Sergej Flach\,\orcidlink{0000-0003-1710-3746}}
    \affiliation{\pcsadd}
    \affiliation{\ustadd}

\author{Alexei Andreanov\,\orcidlink{0000-0002-3033-0452}}
    \email{aalexei@ibs.re.kr}
    \affiliation{\pcsadd}
    \affiliation{\ustadd}

\date{\today}

\begin{abstract}
    We study collective quantum phases and quantum phase transitions occurring in frustrated sawtooth arrays of small quantum Josephson junctions.
    Frustration is introduced through the periodic arrangement of \(0\)- and \(\pi\)- Josephson junctions with the Josephson coupling energies \(\alpha \ej\) of different signs, \(-1\leq \alpha \leq 1\).
    The complexity of the potential landscape of the system is controlled by the frustration parameter \(f=(1-\alpha)/2\).
    The potential energy has a single global minimum in the non-frustrated regime (\(f<\fcr=0.75\)) and a macroscopic number of equal minima in the frustrated regime (\(f>\fcr=0.75\)). 
    We address the coherent quantum regime and identify several collective quantum phases: disordered (insulating) and ordered (superconducting) phases in the non-frustrated regime, as well as highly entangled patterns of vortices and anti-vortices in the frustrated regime. 
    These collective quantum phases are controlled by several physical parameters: the frustration \(f\), the Josephson coupling, and the charging energies of junctions and islands. 
    We map the control parameter phase diagram by characterizing the quantum dynamics of frustrated Josephson junction arrays by spatially and temporally resolved quantum-mechanical correlation function of the local magnetization.
    
    %
\end{abstract}

\maketitle

\section{Introduction}

In recent years there has been continuous growing interest in experimental and theoretical studies of natural~\cite{beloborodov2007granular,goldman2010superconductor} and artificially prepared~\cite{hadley1988phase,rzchowski1997phase,van1996quantum,fazio2001quantum} low dimensional Josephson junction arrays and lattices composed of many small superconducting islands connected by Josephson junctions.
Fascinating phenomena such as the superconductor-insulator quantum phase transition~\cite{goldman2010superconductor,sondhi1997continuous,glazman1997new,haviland2000superconducting}, \(4e\)-condensation in  frustrated Josephson junction diamond chains~\cite{doucot2002pairing,rizzi20064,protopopov2004anomalous}, complex classical and quantum vortex patterns~\cite{van1996quantum,fazio2001quantum}, discrete breathers~\cite{binder2000observation,trias2000discrete} have been observed in such systems, among others. 

Josephson junction arrays have been used for many years as convenient models to study non-linear classical ~\cite{ustinov1998solitons,miroshnichenko2001breathers,andreanov2019resonant} and quantum dynamics~\cite{haviland2000superconducting,matveev2002persistent,seidov2021quantum,pino2016nonergodic} of interacting many-body systems.
Recently superconducting arrays of interacting Josephson junctions were recognized as an excellent experimental platform to realize analogous quantum simulations~\cite{schmidt2013circuit,georgescu2014quantum,acin2018quantum,smith2019simulating}. 
Indeed, a few Josephson junctions embedded in a superconducting loop can form a macroscopic two-level quantum system, i.e., superconducting qubits.
Inductive or capacitive coupling between two such lumped quantum circuits~\cite{krantz2019quantum,kjaergaard2020superconducting,orlando1999superconducting,pernack2024quantum}, or flux quantization induced topological constraints~\cite{orlando1999superconducting,matveev2002persistent,manucharyan2009fluxonium,andreanov2020frustration,neyenhuys2023long} provide tunable long- or short-range interaction in various qubits networks.
Therefore, a variety of interacting  spin-like Hamiltonians can be simulated with Josephson junction arrays. 
On this avenue the seminal Berezinskii–Kosterlitz–Thouless (BKT) transition~\cite{king2018observation}, quantum dynamics of spin-glass patterns~\cite{king2023quantum}, many-body localization~\cite{xu2018emulating}, Majorana edge modes~\cite{mi2022noise},
qubit networks with frustration~\cite{las2014digital,park2022frustrated} and various spin interacting models \cite{smith2019simulating} have been studied using superconducting qubit arrays. 

Another important direction of research of magnetic systems is the study of \emph{frustrated magnets} where the frustration arises either as a result of the competition of ferromagnetic and antiferromagnetic interactions between localized magnetic moments (magnetic frustration)~\cite{anderson1978concept,nisoli2013colloquium} or due to a special geometry of localized magnetic moments arrangement (geometrical frustration)~\cite{anderson1978concept,moessner2006geometrical,schroeder2005competing,balents2010spin,baniodeh2018high,han2012fractionalized}.
Frustration in magnetic systems leads to highly degenerate classical ground states, a large amount of low-lying metastable states and long relaxation times at low temperatures~\cite{moessner2006geometrical,balents2010spin,mahmoudian2015glassy}.
Classical and quantum spin liquids phases have been identified in frustrated magnetic systems~\cite{anderson1978concept,moessner2006geometrical,schroeder2005competing,balents2010spin}. 

The next natural step in the research field of Josephson junctions is then to elaborate on the physics of large \emph{frustrated }Josephson junction arrays and, in particular, to address the coherent quantum regime in which complex collective quantum phases can be obtained.  
While classical non-linear dynamics of such arrays was intensively studied in the past~\cite{rzchowski1997phase,andreanov2019resonant,andreanov2020frustration,caputo2001resonances,valdez2005superconductivity,pop2008measurement}, the quantum regime has received much less attention, see e.g.,~\cite{doucot2002pairing,rizzi20064,protopopov2004anomalous,pernack2024quantum,neyenhuys2023long}.

In this work we present a theoretical study of the macroscopic quantum dynamics occurring in an exemplary quasi-one-dimensional corner-sharing array of Josephson junctions, i.e., a sawtooth frustrated array (see, schematic in Fig.~\ref{fig:schematic}).
In such a system frustration can be conveniently introduced through the periodic arrangement of \(0\)- and \(\pi\)-Josephson junctions, and the frustration parameter \(0 \leq f \leq 1\) is defined as \(f=(1-\alpha)/2\), where \(\alpha\) is the ratio of Josephson couplings energies of \(\pi\)- and \(0\)-Josephson junctions~\cite{andreanov2019resonant,andreanov2020frustration}. 
Experimental realizations of such arrays require \(\pi\)-Josephson junctions that
can be fabricated on the basis of superconductor-ferromagnet-superconductor junctions~\cite{feofanov2010implementation} or various multi-junctions SQUIDS in externally applied magnetic field~\cite{hilgenkamp2008pi,orlando1999superconducting}.
To provide a strong interaction between the Josephson junctions of different cells the capacitance of superconducting islands to the ground is taken into account (see, Fig. ~\ref{fig:schematic}).
As it was shown in Refs.~\cite{andreanov2017resonant,pernack2024quantum} such arrays realize two very different non-frustrated (\(0<f<\fcr \)) and frustrated (\(\fcr<f<1\) regimes, where \(f=\fcr\) is the critical frustration value.
Here, we address both regimes and obtain different collective quantum states (phases) and quantum phase transitions between them. 



The paper is organized as follows: In Section~\ref{sec:model} we introduce the electrodynamic model of frustrated sawtooth quasi-1D arrays of small (\emph{quantum}) Josephson junctions, 
and define the most important physical parameters and dynamic variables of the system. 
We derive the potential and kinetic energies, the total Lagrangian, and identify non-frustrated  (\(f<\fcr\)) and frustrated (\(f>\fcr\)) regimes.
In Section~\ref{sec:qual-description} we present a qualitative description of different phases that can be realized in quantum arrays of Josephson junctions.
In Section~\ref{sec:gen-approach} we describe the general quantitative approach to the calculation of the partition function and the spatio-temporal correlation function.
In Sections~\ref{sec:non-frustrated} and~\ref{sec:frustrated-regime} we study the quantum dynamics of the arrays in non-frustrated and frustrated regimes, accordingly. 
We focus on the numerical and analytical characterization of different collective quantum phases and derive the complete quantum phase diagram \(G-f\), where the so-called \emph{coupling constant} \(G=\sqrt{\ec/\ej}\) controls the quantum dynamics of Josephson junction arrays. 
Section~\ref{sec:conclusion} provides conclusions. 
Appendix~\ref{app:mf} contains details of the simple mean-field approximation used to identify the collective quantum phases in the frustrated regime.

\section{The model and classical nonlinear dynamics}
\label{sec:model}

\begin{figure}
    \includegraphics[width=0.95\columnwidth]{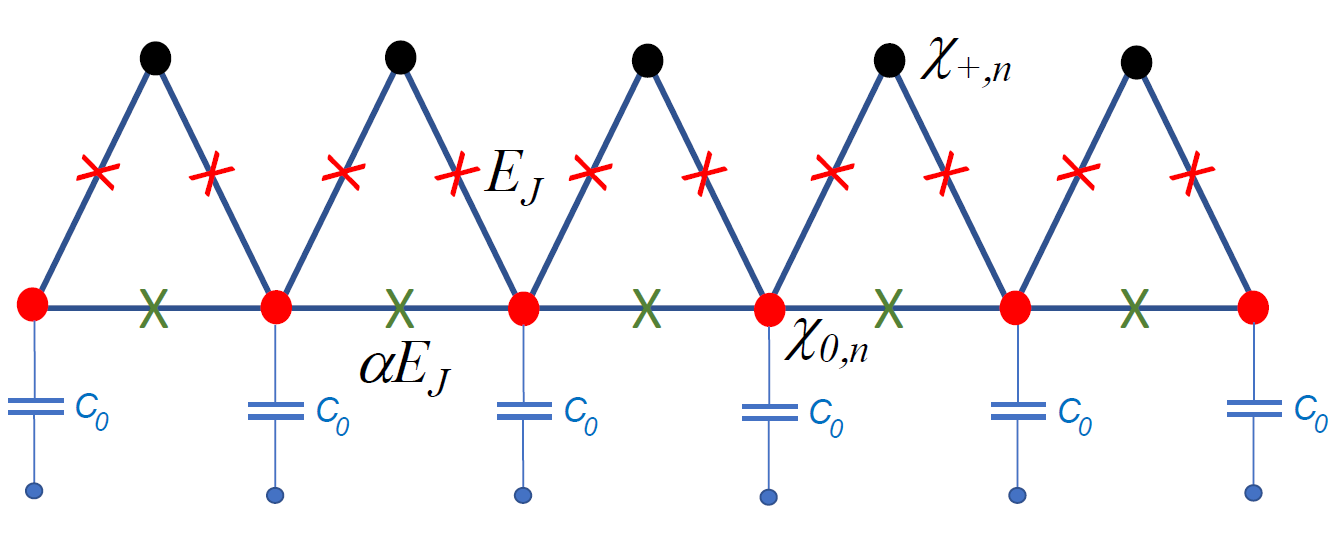}
    \caption{
        Schematics of a frustrated sawtooth  chain of Josephson junctions (indicated by crosses).
        The phases \(\chi_{0,n}\) and \(\chi_{+,n}\) of superconducting islands and the Josephson couplings \(\ej\) and \(\alpha \ej\) in a single cell are shown. 
        \(C_0\) is the capacitance of a superconducting island to the ground.
    }
    \label{fig:schematic}
\end{figure}

We consider a frustrated corner-sharing array of small (quantum) Josephson junctions  where a network of superconducting islands is arranged in a periodic quasi-one-dimensional sawtooth chain with \(N\) cells as shown schematically in Fig.~\ref{fig:schematic}. 
Adjacent superconducting islands are connected by Josephson junctions. 
The Josephson coupling and charging energies in a single cell are fixed at \(\ej\) and \(\ec=4e^2/C\)  (\(C\) is the capacitance of a Josephson junction) for the sides of a triangle, and \(\alpha \ej\) and \(\ec/\abs{\alpha}\) for the base of a triangle, and are repeated periodically in the array. 

We also take into account the charging energy of superconducting islands, \(\ecz = 4e^2/C_0\), where \(C_0\) is the capacitance of a superconducting island to the ground~\cite{pernack2024quantum}.
In most experiments Josephson junctions networks with \(C_0 \ll C\) ( \(\ecz\gg \ec\) ) were used, but the opposite regime \(C_0 \gg C\) was also realized in specially prepared Josephson metamaterials~\cite{ranadive2022kerr,planat2019fabrication}.
In what follows we show explicitly that the charging energies of superconducting islands induce an effective long-range interaction between Josephson junctions of different cells.
This interaction is responsible for the resulting complex collective quantum phases. 
The parameter \(\alpha\) varies in the range \(-1 \leq \alpha \leq 1\), where the so-called \(0\)- and \(\pi\)-Josephson junctions correspond to positive and negative values of \(\alpha\), respectively. 
To quantitatively characterize the properties of the arrays we define the frustration parameter, \(f=(1-\alpha)/2\), which takes values between \(0\) and \(1\)~\cite{andreanov2019resonant,andreanov2020frustration,pernack2024quantum,neyenhuys2023long}.

The classical non-linear dynamics of a quasi-\emph{1D} sawtooth chain of Josephson junctions is completely determined by the \emph{two} time-dependent phases of the superconducting order parameter in each unit cell, \(\chi_n= \{\chi_{0,n}; \chi_{+,n} \}\).
By making use of the Josephson relations we obtain the Lagrangian function \(L=K-U\), where the potential \(U\) and kinetic \(K\) energies of an array are defined as follows
\begin{align}
    U(\{ \chi_n\}) &=  E_J\sum_{n=1}^{N}[2+(1-2f) - \cos (\chi_{+,n}- \chi_{0,n}) - \notag \\
    \label{eq:PotentialEnergy}
    -\cos & (\chi_{+,n} - \chi_{0,n+1})-(1-2f) \cos (\chi_{0,n+1} - \chi_{0,n}) ]
\end{align}
and 
\begin{align}
    K(\{\dot \chi_n\}) &= \frac{\hbar^2}{2E_c}\sum_{n=1}^{N} [(\dot \chi_{0,n} - \dot \chi_{+,n})^2 + (\dot \chi_{0,n+1} - \dot \chi_{+,n})^2 + \notag \\
    \label{eq:KineticEnergy}
    & +|1-2f|(\dot \chi_{0,n+1} - \dot \chi_{0,n})^2 ] +\frac{\hbar^2}{2E_{c0}}\sum_{n=1}^{N+1} (\dot \chi_{0,n})^2.
\end{align}

\begin{figure}
    \includegraphics[width=0.9\columnwidth]{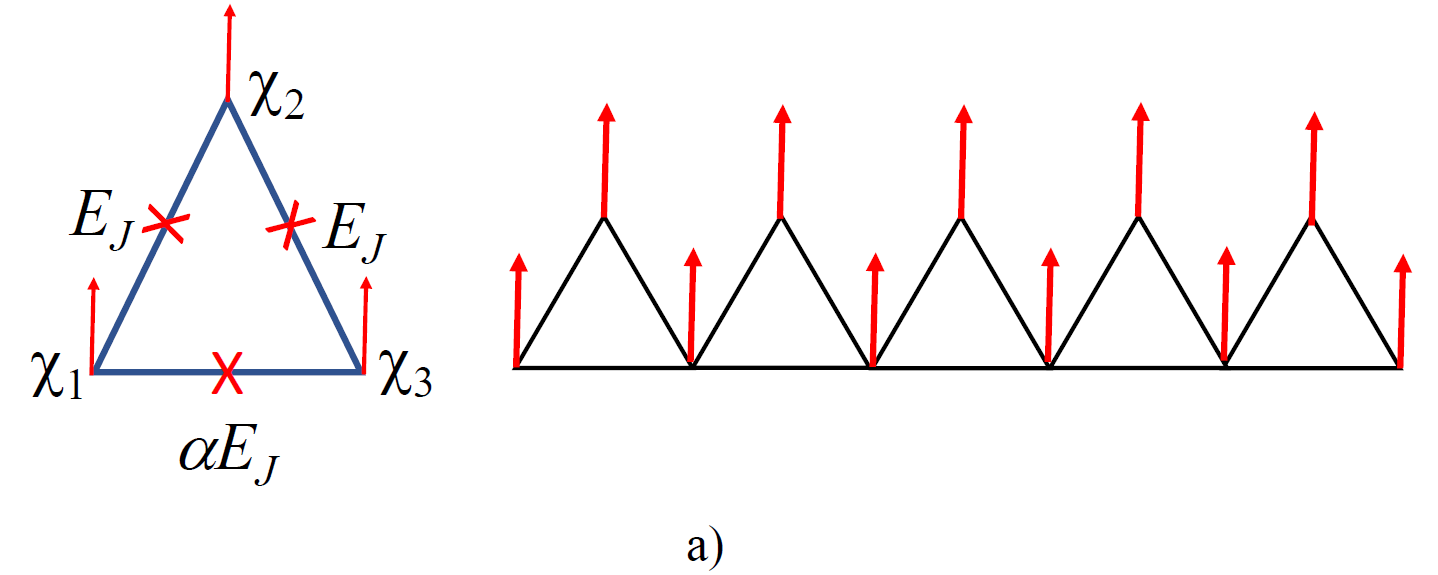}
    \includegraphics[width=0.95\columnwidth]{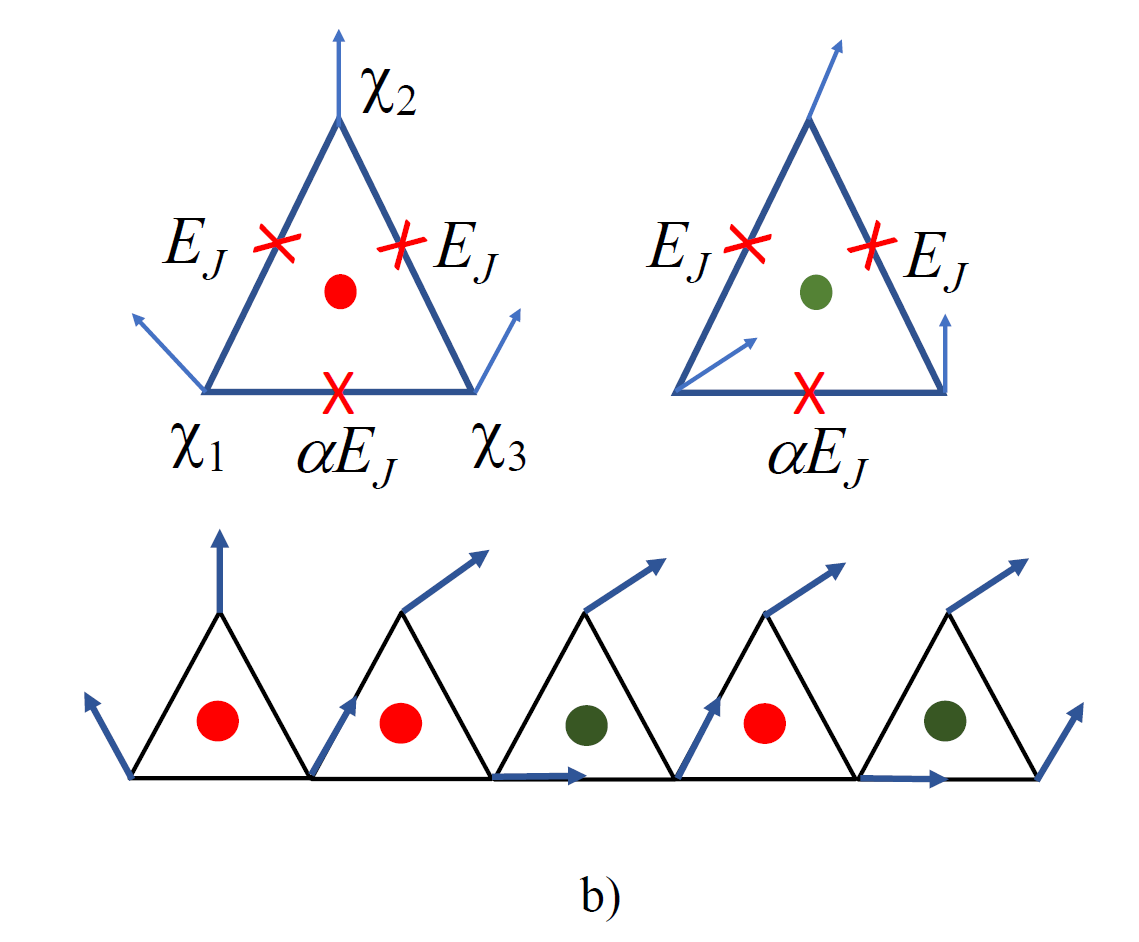}
    \caption{
        Classical ground states of a single triangle (top left and middle) and a sawtooth chain of Josephson junctions (top right and bottom).
        \textit{(a)}: The classical non-frustrated regime (\(f<\fcr\)) is characterized by a single (ferromagnetic) ground state.
        \textit{(b)}: In the frustrated regime (\(f>\fcr\)), each single triangle of Josephson junctions has a doubly degenerated ground state related to the penetration of (anti-)vortices.
        The (anti-)vortices are indicated by (green)red circles, and the phases of the order parameter \(\chi_i\) are shown by arrow.
        A random configuration of vortices-anti-vortices formed in a sawtooth chain of Josephson junctions is shown. 
    }
    \label{fig:Clgroundstates}
\end{figure}

Note that the classical nonlinear dynamics is expected in the limit of \(\ej \gg \ec\) and for moderate temperatures \(T\). 
In Ref.~\onlinecite{andreanov2019resonant} it was shown that there are \emph{two} distinct regimes in this case:
the non-frustrated (\(f<\fcr\)) and the frustrated (\(f>\fcr\)) regimes, with the critical value of frustration \(\fcr=0.75\) separating both.
In the non-frustrated regime the potential energy demonstrates a single global minimum, \(\chi_{+,n} =\chi_{0,n} = 0\) for all \(n\), and, therefore, the ordered ferromagnetic state is realized (see Fig.~\ref{fig:Clgroundstates}(a) for a single cell and for the chain).
As \(f\) approaches the critical value \(\fcr\) one of the two frequencies of small oscillations around the ferromagnetic state vanishes indicating a phase transition, and a transformation of the ground state. 
In the frustrated regime the potential energy landscape contains \(2^N\) equivalent minima determined by the following conditions:
\begin{align}
    \label{eq:GroundStates-1}
    \chi_{+,n} - \chi_{0,n} &= \chi_{0,n} - \chi_{+,n+1}=u(f)/2 \\
    \chi_{0,n+1} - \chi_{0,n} &=u(f) \notag \\
    u(f) &=\pm 2\arccos \left [\frac{1}{4f-2} \right ] \notag.
\end{align}

The minima of the potential energy \(U\) correspond to the penetration of (anti-)vortices in a single cell (see Fig.~\ref{fig:Clgroundstates}(b)), and, therefore, in the absence of interaction between the Josephson junctions of different cells the ground states of the sawtooth array correspond to different sets of vortices and anti-vortices (see Fig.~\ref{fig:Clgroundstates}(d)). 
The vortices and anti-vortices penetrating a single cell correspond to anticlockwise/clockwise flowing persistent currents that are routinely observed in conventional three-junction flux qubits biased at the symmetry point, i.e., as the external magnetic flux \(\Phi \simeq \Phi_0/2\)~\cite{orlando1999superconducting}.

\section{Qualitative description of the quantum dynamics of the sawtooth array}
\label{sec:qual-description}

Next, we provide a qualitative picture of the coherent \emph{quantum dynamics} occurring in the sawtooth array at zero temperature.
The \emph{spatial} properties of the quantum phases depend strongly on the value of \(C_0\).
The presence of even a tiny capacitance to the ground, \(C_0 \ll C\), leads to a \emph{long-range interaction} between well-separated Josephson junctions, and the emergence of a characteristic length scale \(n_0 = \sqrt{C/C_0}\)~\cite{haviland2000superconducting,pernack2024quantum}.

Naturally, one has to distinguish the two limits: \emph{long} (\(N>n_0\)), and \emph{short} (\(N<n_0\)) arrays.
In most artificially fabricated  arrays the characteristic length \(n_0 \gg 1\).
For long arrays the collective quantum phases are expected in both frustrated and non-frustrated regimes.
However for short arrays the interaction between the Josephson junctions in different cells is relatively weak, and the Josephson junctions of different cells display almost independent dynamics with the quantum phase transition lines becoming crossover lines.  

The emergence of a specific quantum collective state is controlled by two parameters: the coupling constant \(G=\sqrt{\ec/\ej}\)~\cite{sondhi1997continuous,haviland2000superconducting} and the frustration \(f\).

For long arrays in the non-frustrated quantum regime, \(f<\fcr\), and for \(G \ll 1\), quantum fluctuations in the form of small amplitude collective oscillations of phases \(\chi_n\) result in non-exponential long-range spatial correlations. 
These correlations correspond to an ordered \emph{superconducting} state (\(LRO-S\)-state), which is well-known in the theory of quantum series arrays of Josephson junctions~\cite{sondhi1997continuous,haviland2000superconducting}.

If the coupling constant \(G\) exceeds a critical value the quantum fluctuations become intensive and the disordered \emph{insulating} state emerges.
These superconducting and insulating states are separated by the quantum phase transition line determined by the condition \(\hbar \onf(f) \simeq E_c\), where \(\onf(f)\) is the frequency of small oscillations around the minimum of \(U\{\chi_n\}\).

In the frustrated regime, \(f>\fcr\), the insulating state is also present for \(G \geq 1\).
However, for \(G \leq 1\) we find instead various collective coherent quantum phases of penetrating vortices and anti-vortices, whose properties depend strongly on the ratio of \(N\) and \(n_0\).

\begin{figure}
    \includegraphics[width=0.95\columnwidth]{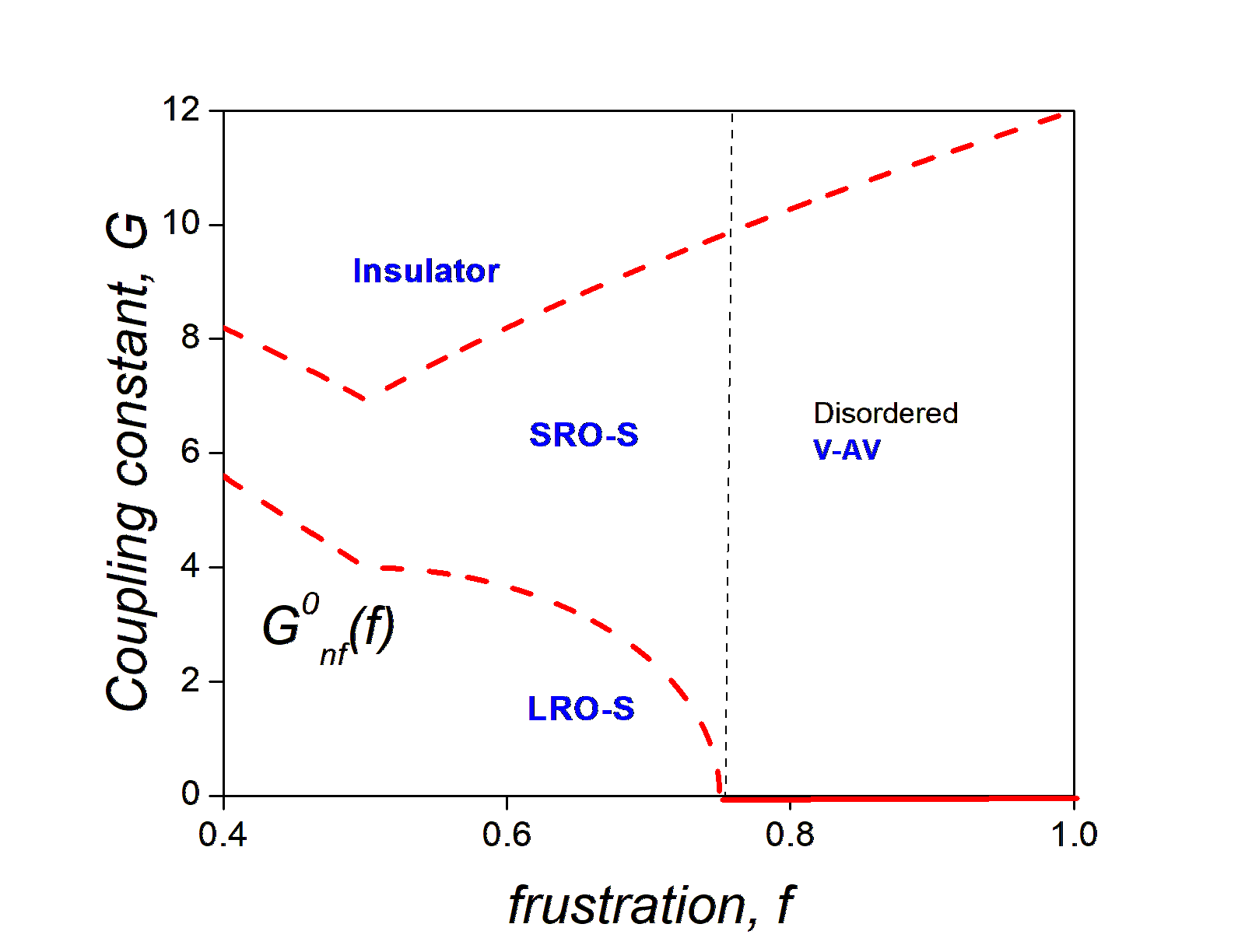}
    \caption{
        Phase diagram of short (\(N<n_0\)) quantum sawtooth chains of Josephson junctions.
        The dashed lines indicate crossovers between different phases, i.e. \(LRO-S\)-, \(SRO-S\)- and insulating states.
    }
    \label{fig:PD-short}
\end{figure}

For \(N<n_0\) individual cells of the array interact weakly with each other, and the quantum-mechanical state of a system factorizes into the direct product of quantum states of individual cells, i.e., \(\ket{\Psi}=\bigotimes_i^N \ket{\psi_i}\).
For \(f>\fcr\) in the frustrated regime, the potential energy \(U\) of a single cell shows two minima corresponding to the penetration of a vortex and an anti-vortex.
The eigenfunction \(\ket{\psi_i}\) consists of the quantum-mechanical superposition of vortex and anti-vortex states. 
Such a superposition of vortex/anti-vortex states corresponds to the coherent quantum-mechanical tunneling between the two minima separated by a barrier. 
The amplitude of tunneling \(\Delta(f)\) is strongly enhanced as the frustration \(f\) approaches the critical value of \(\fcr\). 
We remark here, that in the frustrated regime the quantum superposition of vortex/anti-vortex states in a single superconducting triangle is equivalent to the quantum superposition of persistent currents occurring in a flux qubit biased to the symmetry point~\cite{orlando1999superconducting}. 
The complete phase diagram, \(G-f\), of short arrays is presented in Fig.~\ref{fig:PD-short}.

\begin{figure}
    \includegraphics[width=0.95\columnwidth]{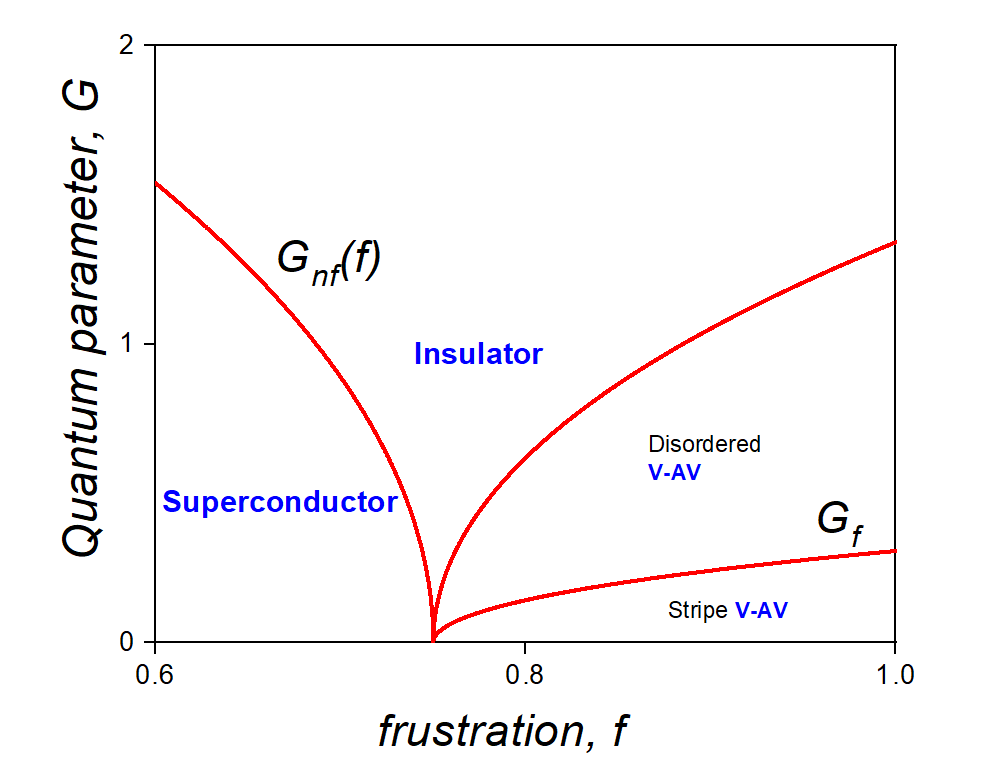}
    \caption{
        Phase diagram of long (\(N>n_0\)) quantum sawtooth chains of Josephson junctions in the presence of a long-range interaction between triangular cells, \(C_0\ll C\).
        Red lines indicate transitions between different quantum phases.
        The parameter \(\beta=0.05\) was chosen.
    }
    \label{fig:PD-long}
\end{figure}

The most interesting case is realized for long arrays with, \(N>n_0\). 
In Section~\ref{sec:frustrated-regime} we show explicitly that the long-range Coulomb interaction produced by Cooper pair charge fluctuations on the islands, induces a quantum phase transition separating two quantum phases.
In the region of \(\gff^{(1)}(f)<G\) the interaction between the cells plays no role and the quantum-mechanical ground state is again a direct  product of states of individual cells, i.e., the superposition of vortex and anti-vortex states with \emph{zero} entanglement. 
In the opposite regime \(G <\gff^{(1)}(f)\) the interaction between vortices and anti-vortices located in well-separated cells leads to a novel collective quantum phase characterized by complex spatio-temporal correlations and high entanglement. 
The \(\gff^{(1)}(f)\) determines the quantum phase transition line separating these quantum phases.
The complete phase diagram, \(G-f\), in the interacting regime as \(N > n_0\), is shown in Fig.~\ref{fig:PD-long}.
Notice here, that the detailed quantitative analysis of the collective quantum phases occurring as \(G \ll \gff^{(1)}(f)\) has been carried out in Ref.~\onlinecite{pernack2024quantum}. 

 






\section{Partition function and spatio-temporal correlation functions of frustrated arrays}
\label{sec:gen-approach}

The analysis of the quantum regime of the sawtooth array can be greatly simplified close to the critical frustration, \(\fcr\), as the variables \(\chi_{+,n}\) perform small oscillations with a frequency larger than the ones of \(\chi_{0,n}\)~\cite{andreanov2019resonant}.
At low temperatures these high frequency oscillations cannot be excited and, thus, for low-lying quantum states close to the ground state the phases \(\chi_{+,n}\) can be treated as classical variables: \(\chi_{+,n}=(\chi_{0,n}+\chi_{0,n+1})/2\).
Substituting this expression in Eqs.~(\ref{eq:PotentialEnergy}-\ref{eq:KineticEnergy}) we rewrite the effective potential and kinetic energies in the following form:
\begin{align}
    \label{eq:PotentialEnergyeff}
    \ueff(\{ \chi_{0,n}\}) &=  E_J\sum_{n=1}^{N} 
    \left \{ 3-2f- 2\cos [(\chi_{0,n+1}- \chi_{0,n})/2] \right \} - \notag \\
    & -(1-2f)E_J\sum_{n=1}^{N} [1- \cos (\chi_{0,n+1} - \chi_{0,n})] 
\end{align}
and 
\begin{align}
    \label{eq:KineticEnergyeff}
    \keff(\{\dot \chi_{0,n} \}) &= \frac{\hbar^2 (1+2|1-2f|)}{4E_c} \sum_{n=1}^{N}  [(\dot \chi_{0,n} - \dot \chi_{0,n+1})^2  + \notag \\
    & +\frac{\hbar^2}{2\ecz}\sum_{n=1}^{N+1} (\dot \chi_{0,n})^2.
\end{align}
The different quantum collective phases occurring in  the arrays can be quantitatively characterized through the calculation of the partition function \(Z\) and the spatio-temporal quantum-mechanical correlation function of the local \emph{magnetization} \(\BC(n,t)=\langle\cos[\chi_{0,0}(0)-\chi_{0,n}(t)] \rangle\), where \(\langle\dots\rangle\) is the quantum-mechanical averaging at zero temperature.

The partition function \(Z\) is expressed as a path integral in the imaginary time-representation:
\begin{gather}
    \label{eq:PartFunction}
    \BZ = \int D[\chi_{0,n}(\tau_1)] \exp \left [-\frac{1}{\hbar}\int_0^{\kbt}\heff\{\chi_{0,n}, \dot \chi_{0,n} \} d \tau_1 \right],
\end{gather}
where 
\begin{gather}
    \label{eq:HamiltonianEff}
    \heff = \keff + \ueff
\end{gather}
The correlation function \(\BC(n,\tau)\) is then
\begin{gather}
    \BC(n,\tau) = \frac{1}{Z}\int D[\chi_{0,n}(\tau_1)]\cos[(\chi_{0,0} (0) - \chi_{0,n}(\tau)] \times \notag \\
    \times \exp \left [\frac{1}{\hbar}\int_0^{\kbt}\heff\{\chi_{0,n}, \dot \chi_{0,n}  \} d \tau_1 \right].
    \label{eq:CorrFunction}
\end{gather}
The analytical continuation of \(\BC(n,\tau)\) to the real axis gives \(\BC(n,t)\).

\section{Collective Quantum phases in the non-frustrated regime \(f<\fcr\)}
\label{sec:non-frustrated}

In the non-frustrated regime, \(f<\fcr=0.75\), the effective potential energy \(\ueff\)~\eqref{eq:PotentialEnergyeff} has a single global minimum with all \(\chi_{0,n}=0\).
The coherent quantum dynamics in this regime is determined by small amplitude collective oscillations around the position of a global minimum.
Expanding \(\ueff\) around the minimum up to the second order, the effective Hamiltonian \(\heff\) becomes (\(u_n=\chi_{0,n+1}-\chi_{0,n}\))
\begin{align}
    \label{eq:PotentialEnergyExpending}
    \heff &=\sum_{n=1}^{N} \frac{\hbar^2 (1+2|1-2f|)}{4E_c} \left [\frac{du_n}{d\tau} \right ]^2+\frac{E_J(3-2f)}{2} \frac{u^2_n}{2}+ \notag \\
    & +\frac{\hbar^2}{2\ecz}\sum_{n=1}^{N} \left (\frac{d\chi_{0,n}}{d\tau} \right )^2.
\end{align}
Note that this approximation is not valid for \(G \gg 1\), i.e., deeply in the insulating regime.  
Within this approximation the Hamiltonian \(\heff\) is quadratic: evaluating all the Gaussian integrals over \(\chi_{0,n}\) in Eqs.~\eqref{eq:PartFunction} and~\eqref{eq:CorrFunction} we obtain the \emph{spin-wave} contribution to the correlation function \(\BC(n,\tau)\):
\begin{align}
    \label{eq:Correlationfunction2}
    \BC(n,\tau)=\exp \left [ \frac{E_c}{2\hbar \gamma} \int \frac{dx dy}{(2\pi)^2} \frac{\exp\{ ixn+iy\tau\}-1}{y^2( x^2+2\beta/\gamma)+\onf^2 x^2} \right],
\end{align}
where \(\onf=\sqrt{E_J E_c(3-2f)/(\gamma\hbar^2)}\) is the frequency of small amplitude collective oscillations in the non-frustrated regime; 
\(\beta=E_c/\ecz\) is the ratio of the Josephson charging energy to the charging energy of a superconducting island, and \(\gamma=1+2|1-2f|\) is of order one.
Evaluating the integral over \(y\) in Eq.~\eqref{eq:Correlationfunction2} we obtain
\begin{align}
    \label{eq:Correlationfunction3}
    \BC(n,\tau)=\exp \left [ \frac{E_c}{4\hbar\onf \gamma} \int \frac{dx}{2\pi} \frac{\exp \left \{ ixn-\frac{|x|\onf|\tau|}{\sqrt{ x^2 + 2\beta/\gamma}} \right \}-1}{|x|\sqrt{x^2 + 2\beta/\gamma}} \right].
\end{align}
%
The spatio-temporal behavior of the correlation function \(\BC(n,\tau)\) can be evaluated in two limits as discussed below.
For \emph{long} arrays, \(N \gg n_0=\sqrt{\gamma/(2\beta)}\), the integral over \(x\) in Eq.~\eqref{eq:Correlationfunction3} has a logarithmic singularity.
Therefore, the correlation function at large distances \(n\), i.e., \(n \gg n_0 \), or large times, \(t \gg t_0=1/\onf\), exhibits a power-law decay
\begin{align}
    \label{eq:Correlationfunction_Cneq0}
    \BC(n,\tau)= \left [\frac{n_0}{\max \left\{n; \onf\sqrt{\frac{\gamma}{2\beta}}\tau \right \} }\right ]^g,
\end{align}
with the exponent \(g=E_c/(8\sqrt{2}\pi \hbar \onf\sqrt{\beta \gamma})\). 
Such a power-law  behavior indicates the appearance of a collective superconducting phase. 
It is well-known from the study of the quantum phase transition in series arrays of Josephson junctions~\cite{sondhi1997continuous} and the general Berezinskii-Kosterlitz-Thouless transition~\cite{kosterlitz1973ordering} that the quantum phase transition between the \emph{superconducting} and the \emph{insulating }states occurs at \(g=1/4\), thus fixing the quantum phase transition line:
\begin{align}
    \label{eq:phase_transition_lineC-NF}
    G=G_\mathrm{nf}=2\pi \sqrt{2\beta(3-2f)}.
\end{align}
These collective quantum phases are presented on the phase diagram shown in Fig.~\ref{fig:PD-long}.

In the opposite limit of \emph{short} arrays, \(N \ll n_0\), the integral in Eq.~\eqref{eq:Correlationfunction3} can be simplified to
\begin{align}
    \label{eq:Correlationfunction_C=0}
    \BC(n,\tau)=\exp \left [ \frac{E_c}{4\hbar\onf \gamma} \int \frac{dx}{2\pi} \frac{\exp \left \{ixn - \onf\abs{\tau} \right \}-1}{x^2 } \right].
\end{align}
Evaluating the integral over \(x\), we find that for zero time lag (\(\tau=0\)) the correlation function demonstrates exponentially decaying correlations in space
\begin{align}
    \label{eQ:Correlationfunction_C=0_SpCorr}
    \BC(n,0)=\exp \left [ -\frac{E_c}{4\hbar\gamma \onf}|n| \right].
\end{align}
Therefore, the line
\begin{align}
    \label{eQ:phaseline-nf0}
    G=G^0_\mathrm{nf}(f) = 4\sqrt{\gamma(3-2f)}
\end{align}
describes the crossover between the superconducting states with long- (low \(G\) values) and short- (high \(G\) values) spatial correlations.
The correlation function \(\BC(0,\tau)\) decays exponentially with \(\tau\), corresponding to the collective oscillations of frequency \(\onf\) in these \(LRO\)- and \(SRO\)- \emph{superconducting} states. 
However, as \(G \geq 4\sqrt{3}\sqrt{1+2|1-2f|}\), the intensive \(2\pi\)-phase slips occur, and the \emph{insulating} state emerges.
These quantum states are presented on the phase diagram shown in Fig.~\ref{fig:PD-short}. 

To conclude this section we remark that the characteristic length is rather large, \(n_0 \gg 1\), in the limit of \(C_0 \ll C\).


\section{Collective Quantum phases in the frustrated regime \(f>\fcr\)}
\label{sec:frustrated-regime}


In the frustrated regime (\(f>\fcr=0.75\)), the effective potential of a single cell acquires a double well potential form, that we exploit in our derivation.
Our approximation is inspired by a method that was proposed long time ago to study hydrogen-bonds vibrations in the ferroelectrics~\cite{blinc1960isotopic}.
The effective potential energy \(\ueff\)~\eqref{eq:PotentialEnergyeff} can be presented as a sum of independent terms, i.e., \(\ueff=\sum_n \ueff(u_n)\), and each term has two shallow minima separated by a barrier, (\(\ueff(u_n)=E_J [2+\alpha-2\cos(u_n/2)-\alpha \cos u_n ]\)).
We approximate the potential \(\ueff(u_n)\) as
\begin{align}
    \label{eq:PotEnergy:fr}
    \tilde{U}_{\mathrm{eff}}(u_n, s_n)=U_{\mathrm{min}}+\tilde{E}_J \left [\frac{u_n^2}{2a^2} - s_n \frac{u_n}{a} \right],
\end{align}
where \(\tilde{E}_J=E_J[(4f-3)^2]/(2f-1)\), \(a=2 \acos [1/(4f-2)]\).
The positions of the minima and the barrier height match the ones for the original potential \(\ueff(u_n)\). 
The \(s_n=\pm 1\) are the classical Ising "spins": values \(\mp 1\) correspond to the (anti-)vortex penetrating each cell. 
This approximation allows one to represent the total Hamiltonian \(\heff\{\chi_{0,n}, \dot \chi_{0,n} \}\) as a quadratic function over all \(\chi_{0,n}\) variables.

 

Using the Suzuki-Trotter decomposition~\cite{trotter1959product} and discretizing the total imaginary time interval \([0,\kbt]\) with \(M\) points \(\tau_m\), where \(m=1...M\), we calculate the Gaussian integrals over \(\chi_{0,n}(\tau_m)\) in the Eq.~\eqref{eq:PartFunction}, and express the partition function \(Z\) as a classical partition function of the Ising "spins" \(s_n\) 
\begin{align}
    \label{eq:Partitionfunction-FrRegime}
    Z=\Tr\{s_{\vec{r}}\}\exp \left [ -\frac{ \tilde{E}_J^2 (\delta \tau)^2 \sqrt{E_c/\tilde{E}_J}}{ 2\sqrt{2\gamma} \hbar^2 a}\sum_{\vec{r}_1,\vec{r}_2} s_{\vec{r}_1}s_{\vec{r}_2}\Gamma (\vec{r}_1-\vec{r}_2)\right],
\end{align}
where \(\vec r\) is the two-dimensional vector of the square lattice \(\{n,m\}\).
Here, the sets of integers \(n\) and \(m\) determine the spatial and temporal directions, accordingly. 
The Green function on the two-dimensional square lattice, \(\Gamma (\vec{r}_1-\vec{r}_2)\), determines the interaction between two classical spins, \(\pm 1\), located at \(\vec{r}_1\) and \(\vec{r}_2\).
\(\Gamma (\vec{r})\) is given by
\begin{align}
    \label{eq:Greenfunction}
    \Gamma (\vec{r}=\{n,m\}) = \frac{d^2}{dn^2} \int \frac{dx}{2\pi}  \left [\frac{\exp \left \{ ixn-|m|\frac{ |x|\omega_\mathrm{f} \delta \tau}{\sqrt{ x^2+2\beta/\gamma}} \right \}}{|x|\sqrt{ x^2+2\beta/\gamma}} \right ],
\end{align}
where \(\omega_\mathrm{f}=(1/a)\sqrt{2E_c\tilde{E}_J/(\gamma \hbar^2)}\) and \(\delta \tau=\tau_{m+1}-\tau_m\).

Thus, in the frustration regime the coherent quantum dynamics of sawtooth arrays is reduced to an Ising Hamiltonian of \emph{interacting classical spins} \(s_n(\tau_m)\) on the two-dimensional square lattice \(\{n,m\}\).

We analyze the partition function~(\ref{eq:Partitionfunction-FrRegime},~\ref{eq:Greenfunction}) in two limits: short (\(N\ll n_0\)) and long (\(N \gg n_0\)) sawtooth chains of Josephson junctions.

\subsection{Short arrays, \(N \ll n_0\): spatio-temporal disordered quantum states of vortices and anti-vortices.}
\label{subsec:1D-Ising}

For short  arrays the spatial dependence of \(\Gamma (n,m)\) is reduced to a \(\delta\)-function, and therefore, in this limit the vortices and anti-vortices penetrating different cells do not interact with each other. 
In the temporal direction, \(m\), the Green function is short-ranged, and decays exponentially over the distance \(\simeq (\omega_f \delta \tau)^{-1}\).
Therefore, choosing \(\delta \tau =1/\omega_f\) the original two-dimensional problem of interacting spins can be reduced to a one-dimensional array of spins with the nearest-neighbor interaction in the temporal direction:
\begin{align}
    \label{eq:Partitionfunction-FrRegime0}
    Z=\Tr\{s_m\}\exp \left [ K_0\sum_{m=1}^M s_{m}s_{m+1}\right].
\end{align}
Here a single parameter \(K_0 \simeq [\sqrt{\gamma}/(4\sqrt 2)]a \sqrt{\tilde{E}_J/E_c}\) was introduced, and the total number of grids is \(M=\hbar \omega_\mathrm{f}/(k_B T) \gg 1\).

The partition function~\eqref{eq:Partitionfunction-FrRegime0} corresponds to the 1D Ising model with nearest-neighbor ferromagnetic interaction. 
It is well-known that this model does not exhibit any quantum phase transition, and remains disordered for arbitrary values of \(K_0\).
The partition function \(Z\) and the temporal correlation function \(\BC (\tau)\) can be evaluated by the transfer-matrix method~\cite{baxter2016exactly,strecka2015brief,suzuki1985transfer}, giving \(\BC(\tau)=\exp (-\Delta \tau/\hbar)\) with \(\Delta \simeq \hbar \omega_{\mathrm{f}}\exp (-K_0)\). 
\(\Delta\) is the energy gap between the ground state and the first excited state. 
Analytic continuation of \(\BC(\tau)\) to the real time axis leads to \(\BC(t)\) oscillating in time with the frequency of \(\Delta/\hbar\).
This frequency also determines the frequency of vortex/anti-vortex quantum beats in a single superconducting triangle. 

Thus, one can conclude that for short arrays the interaction between Josephson junctions of different triangles is weak, and the ground state of the array is a direct product of quantum states of individual cells that present symmetric quantum superposition of vortex and anti-vortex states,
i.e., a \emph{spatially disordered vortex/anti-vortex} quantum phase is established. 
Such a quantum state has zero entanglement, and the quantum beats occurring in different triangles are not correlated.  

The crossover line between such a state and the insulating quantum phase  is determined by the condition, \(G = 4\sqrt{3}\sqrt{1+2|\alpha|}\) where intensive \(2\pi\)-phase slips occur. 
The complete phase diagram of \(G-f\) for \(N \ll n_0\) is presented in Fig.~\ref{fig:PD-short}.

\subsection{Long arrays, \(N \gg n_0\): spatio-temporal stripe quantum states of vortices and anti-vortices} 
\label{subsec:2D-Ising}

For long sawtooth arrays, \(N \gg n_0\), the integral over \(x\) in Eq.~\eqref{eq:Greenfunction} has a logarithmic singularity, and therefore, the partition function is written as:
\begin{align}
    \label{eq:Partitionfunction-FRC0}
    Z &= \Tr_{\{s_{\vec{r}}\}} \exp [ K \sum_{\vec{r}_1,\vec{r}_2} s_{\vec{r}_1}s_{\vec{r}_2}f(\vec{r}_1-\vec{r}_2)], \\
    K &= [\sqrt{\beta}/(4\pi a)] \sqrt{\tilde{E}_J/E_c}, \notag \\
    \label{eq:Greenfunction_FRC0}
    f(\vec{r}) &= -\cos (2\theta)/r^2.
\end{align}
Here, \(\vec{r} = \vec{r}_1 - \vec{r}_2\) and \(\theta\) is the polar angle of the vector \(\vec{r}\) on the two-dimensional lattice \(\{n,m\}\). 
In this regime the low energy quantum-mechanical dynamics of the array is determined by a 2D classical Ising spin model with long-range interaction \(K f(\vec{r})\). 
This interaction is ferromagnetic along the \(y\) ("imaginary time") direction, and antiferromagnetic along the \(x\)-direction. 

The collective phases occurring in this model are well identified within the mean-field approximation due to the long-range character of the interaction.
Introducing local auxiliary magnetic fields \(h^\star_{\vx}\) we obtain the transcendent mean-field equation (see details in Appendix~\ref{app:mf}):
\begin{gather}
    \label{eq:mf}
    h^\star_{\vx} = \sqrt{2K}\sum_{\vy} f(\vx - \vy)\tanh(\sqrt{2K} h^\star _{\vy}),
\end{gather}
and the averaged local magnetization, \(m_{\vx}=\langle s_{\vx}\rangle\) is determined as,
\begin{gather}
    \label{eq:localmagn}
    m_{\vx} = \tanh(\sqrt{2K} h^\star_{\vx}).
\end{gather}
For low values, \(K < \kcr\), the unique solution of Eq.~\eqref{eq:mf} is \(h^\star_{\vx}=0\), corresponding to the \emph{disordered} spin phase with the zero local magnetization \(m_{\vx}\). 
However, for \(K > \kcr\) this solution becomes unstable, and the model exhibits a phase transition into the \emph{stripe} spin phase, the local magnetization has a constant value along the \(y\)-direction but changes sign on adjacent sites along the \(x\)-direction.

The mean-field solution of Eq.~\eqref{eq:mf} for auxiliary magnetic fields has a form \(h^\star (x)=(-1)^x h\),
where \(x\) is the spatial coordinate of the stripe, and, therefore, 
\begin{figure}
    \includegraphics[width=0.95\columnwidth]{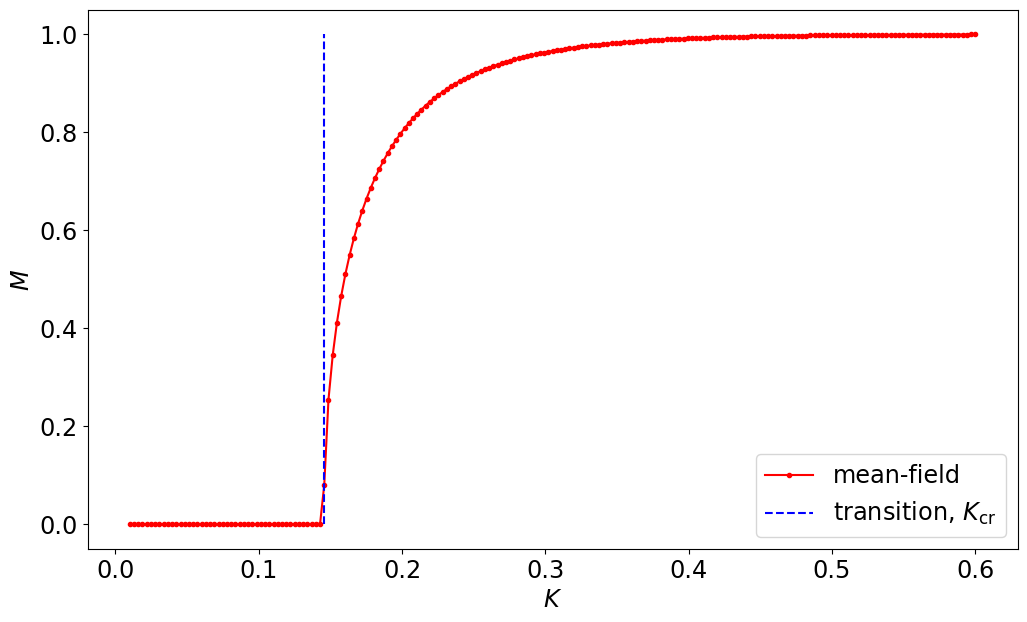}
    \caption{
        The dependence of the local magnetization inside of a single stripe on the parameter \(K\) within the mean-field approximation of the effective model~\eqref{eq:Partitionfunction-FRC0}.
        The vertical dashed line indicates the mean-field prediction for the transition between disordered and stripes spins phases. 
    }
   \label{fig:gs-fru}
\end{figure}
\begin{figure*}[t]
\begin{center}
    \includegraphics[width=220pt]{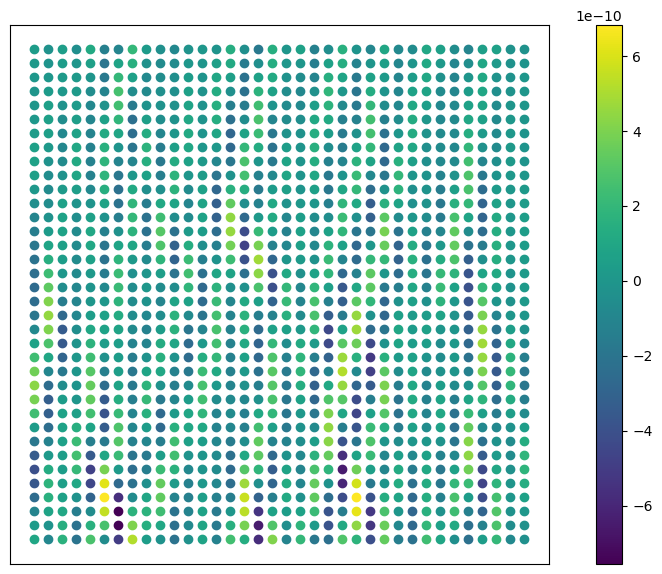}
    \includegraphics[width=220pt]{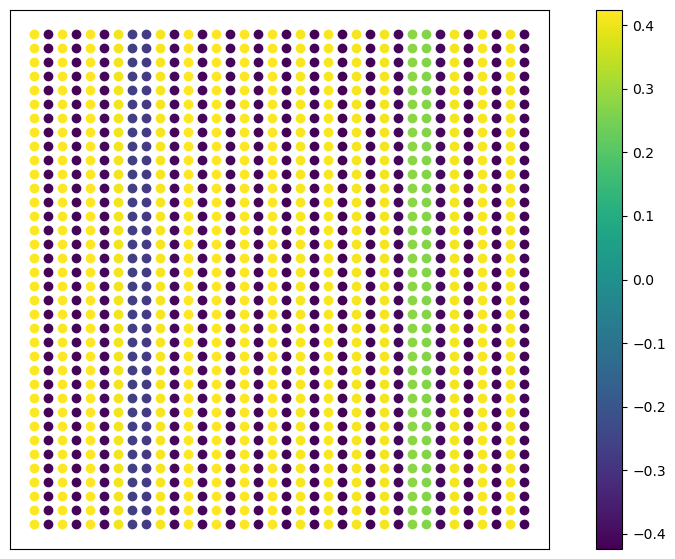}
    \includegraphics[width=220pt]{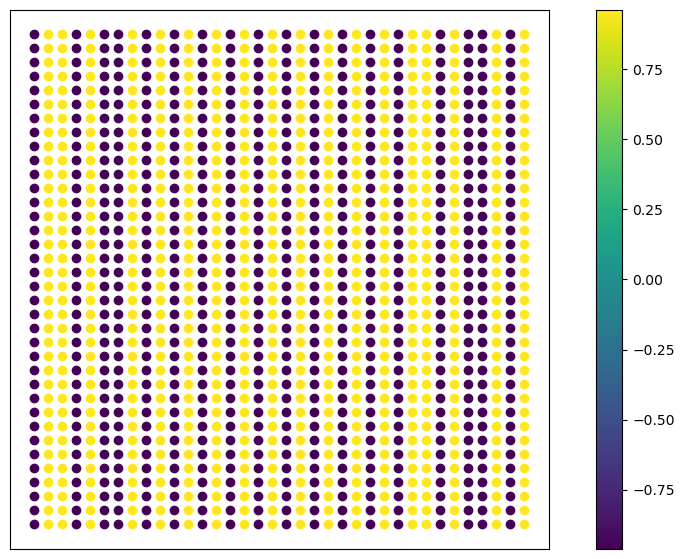}
    \caption{
        Quenches from \(K=0\) within the mean-field approximation of the effective model~\eqref{eq:Partitionfunction-FRC0}.
        The plots show mean-field onsite local magnetization patterns for several values of \(K\): 
        low, \(K=0.0912\) (top left), close to criticality, \(K=0.155\) (top right), high, \(K=0.2942\) (bottom).
    }
   \label{fig:gs-fru-2}
\end{center}
\end{figure*}
solving Eqs.~\eqref{eq:mf} and~\eqref{eq:localmagn} numerically we obtain the dependence of the averaged local magnetization on the parameter \(K\) clearly showing this phase transition, Fig~\ref{fig:gs-fru}. 
The critical value of \(\kcr = 1/(2\lambda_{\mathrm{max}}) \simeq 0.16\) is determined by the largest eigenvalue \(\lambda_{\mathrm{max}}\) of the linearized operator in r.h.s. of Eq.~\eqref{eq:mf}, i.e., \(\hat \BG h^\star _{\vx}=\lambda_{\mathrm{max}}h^\star_{\vx}\), where \(\hat \BG=\sum_{\vy} f(\vx - \vy) h^\star _{\vy}\). 
Typical local magnetization patterns obtained for low, close to critical, and high values of \(K\) are shown in Fig.~\ref{fig:gs-fru-2}.


    
Turning back to the description of the frustrated regime in the framework of penetrating vortices and anti-vortices and their quantum superposition, we identify two collective quantum states that can be observed. 
The first one is the already elaborated in Sec.~\ref{subsec:1D-Ising}: spatially disordered quantum superposition of vortex-anti-vortex states corresponding to a spatially-temporal disordered spin state of the Ising  model~\eqref{eq:Partitionfunction-FRC0} for \(K < \kcr\).
The second one, i.e., the spatially-temporal stripe spin phase of the Ising spin model~\eqref{eq:Partitionfunction-FRC0}, is the collective vortex/anti-vortex state characterized by the ferromagnetic ordering in "temporal" direction that corresponds to a strong suppression of the quantum superposition of a vortex and anti-vortex in a single cell.
In this state alternated regions of vortex(anti-vortex) of a width, \(n_0 \simeq \sqrt{\gamma/(2\beta)}\), are found. 
The quantum phase transition line between two states is given by
\begin{gather} 
    \label{eq:G-f}
    G = G_\mathrm{f} = \frac{1}{4\pi a}\sqrt{\frac{\beta}{\abs{\alpha}}} (2\abs{\alpha} - 1)
\end{gather}
To conclude this section we notice that the quantum phase transition line between the disordered \(V\)-\(AV\) state and the insulator state is determined by the condition \(E_c/(2\pi \sqrt{2}\hbar \omega_f \sqrt{\beta \gamma})=1\) driven by the intensive collective quantum mechanical fluctuations. 
The complete phase diagram \(G\)-\(f\) as \(C_0 \ll C\) is presented in Fig.~\ref{fig:PD-long}.

\section{Conclusions}
\label{sec:conclusion}

In conclusion we developed a systematic theoretical study of collective quantum states occurring in one-dimensional sawtooth frustrated Josephson junction arrays. 
In such superconducting quantum circuits the frustration is introduced as the periodic alternated arrangement of \(0\)- and \(\pi\)- Josephson junctions.
The parameter of frustration \(f\) is determined by the strength of the \(\pi\)-Josephson junction. 
The \emph{classical} sawtooth array demonstrates two very different regimes: the non-frustrated regime, \(f<\fcr\), is characterized by the unique classical ground state, and in the frustrated regime, (\(f>\fcr\)), the highly degenerated ground state is realized in the form of different sets of vortex (\(V\)) and anti-vortex (\(AV\)) penetrating an each superconducting cell.
These \textit{V} (\textit{AV}) states correspond to the (anti-)clockwise persistent currents flowing in a single superconducting triangle~\cite{orlando1999superconducting,neyenhuys2023long,pernack2024quantum}. 

As we turn to the coherent quantum dynamics of sawtooth arrays, the long-range interaction between Josephson junctions of different cells provided by the capacitance \(C_0\) to the ground, determines the observed collective quantum phases.
Moreover, we distinguish two cases, i.e., long and short arrays, depending on the ratio of \(N\) and \(n_0\), where the length \(n_0\) is a known charge screening length~\cite{haviland2000superconducting,pernack2024quantum}.
In the non-frustrated quantum regime we identified the superconducting and insulating states depending on the values of parameters \(G=\sqrt{E_c/E_J}\) and \(f\).
For long arrays the quantum phase transition between these phases is of the BKT type~\cite{sondhi1997continuous,haviland2000superconducting,kosterlitz1973ordering}.

In the frustrated quantum regime the ground state of the basic element of an array, i.e., a single superconducting triangle consisting of two \(0\)- and one \(\pi\)-Josephson junctions, is the symmetric quantum superposition of the \(V\) and \(AV\) eigenstates.
Notice here, that the symmetric and antisymmetric quantum superposition of the \textit{V/AV} eigenstates determine the macroscopic quantum two-levels system corresponding to the \emph{flux qubit} biased to the symmetry point~\cite{orlando1999superconducting}.
For short arrays the interaction between the Josephson junctions of different cells is weak, and therefore, the quantum dynamics represents the one of the independent flux qubits.
The complete phase diagram \(f-G\) for short arrays, \(N<n_0\), is presented in Fig.~\ref{fig:PD-short}.

The most interesting results are found for long arrays, \(N>n_0\), in which the capacitance to the ground, \(C_0 \neq 0\), generates a long-range interaction between Josephson junctions (vortices and anti-vortices penetrating different cells).
In the frustrated quantum regime using a mapping from the quantum circuit Hamiltonian (\ref{eq:PotentialEnergyeff} and \ref{eq:KineticEnergyeff}) to the effective 2D (space-imaginary time) long-range Ising model (see, Eq.~\eqref{eq:Partitionfunction-FrRegime}) supplemented by the mean-field analysis, we identified two distinguished quantum phases of penetrating \textit{V}s/\textit{AV}s.
For moderate values of \(G\) we obtained the spatial-temporal disordered state of \textit{V}s and \textit{AV}s, which is characterized by low entanglement, and large amplitude of tunneling between \textit{V} and \textit{AV} states of a single cell (see, top left of Fig.~\ref{fig:gs-fru-2}).

For small values of \(G\) we obtained the \emph{stripe} quantum state of \(V\)s and \(AV\)s ((see bottom of Fig.~\ref{fig:gs-fru-2}.).
Since the stripes are stretched along the temporal direction, the quantum superposition of \(V\) and \(AV\) in a single cell is strongly suppressed, and a spatially alternating state of \(V\) and \(AV\) is established.
The complete phase diagram \(f-G\) for long sawtooth chains of Josephson junctions is presented in Fig.~\ref{fig:PD-long}.
The line of the quantum phase transition between the disordered and stripes collective phases of \(V\) and \(AV\)s was obtained in the mean-field approximation of Eq.~\eqref{eq:Partitionfunction-FRC0}.

Finally we notice that the detailed characterization of the stripe phase in the limit of \(G\ll G_f\), i.e., not too close to the transition line, has been carried out in Ref.~\onlinecite{pernack2024quantum} by the variational method.

\begin{acknowledgments}
    We acknowledge the financial support through the European Union’s Horizon 2020 research and innovation program under grant agreement No 863313 'Supergalax'.
    AA and SF acknowledge support of the Institute for Basic Science (IBS) in the Republic of Korea through the project No. IBS-R024-D1.  
\end{acknowledgments}

\appendix

\section{Classical Ising spins with long-range interaction: mean-field analysis}
\label{app:mf}

The partition function \(Z\) of classical Ising spins determined by Eqs.~\eqref{eq:Partitionfunction-FrRegime} and~\eqref{eq:Greenfunction} is quantitatively analyzed in the framework of the mean-field theory. 
For that we adapted a slightly modified version of the standard Curie-Weiss mean-field theory~\cite{strecka2015brief,suzuki1985transfer}. 
Introducing auxiliary magnetic fields \(h_{\vx}\) and calculating all sums over spins \(s_{\vx}\) we obtain
\begin{gather}
    Z= \Tr \{s_{\vec{r}}\} \exp\left[K\sum_{\vx\neq \vy} f(\vx, \vy) s_{\vx} s_{\vy}\right] = \\
    \int\prod_{\vx} dh_{\vx} \Tr \{s_{\vec{r}}\} \exp\left[-\frac{1}{2}\sum_{\vx,\vy} h_{\vx} f_{\vx,\vy}^{-1} h_{\vy} +\sqrt{2K} \sum_{\vx} h_{\vx} s_{\vx} \right] \\
    = \int\prod_{\vx} dh_{\vx} \exp\left[-\frac{1}{2}\sum_{\vx,\vy} h_{\vx} f_{\vx,\vy}^{-1} h_{\vy} + \sum_{\vx} \ln\cosh(\sqrt{2K} h_{\vx})\right].
\end{gather}
In the mean-field approximation the partition function \(Z\) is determined by the saddle point solution, \(h^\star _{\vx}\):
\begin{gather}
    -f_{\vx,\vy}^{-1} h^\star_{\vy} + \sqrt{2K}\tanh(\sqrt{2K} h^\star_{\vx}) = 0\
\end{gather}
or
\begin{gather}
    \label{eq:mf-1}
    h^\star_{\vx} = \sqrt{2K}\sum_{\vy} f(\vx - \vy)\tanh(\sqrt{2K} h^\star _{\vy}).
\end{gather}
The local magnetization is determined by auxiliary magnetic fields \(h^\star_{\vx}\) as
\begin{gather}
    m_{\vx} = \tanh(\sqrt{2K} h^\star_{\vx}).
\end{gather}

We define here the free energy as:
\begin{gather}
    F_\mathrm{MF}= \frac{1}{2K} \sum_{\vx,\vy} h^\star_{\vx} f_{\vx,\vy}^{-1} h^\star_{\vy} - \frac{1}{K}\sum_{\vx} \ln\cosh(h^\star_{\vx} \sqrt{2K})= \\
     = \frac{1}{K} \sum_{\vx} \left[\sqrt{\frac{K}{2}}h^\star_{\vx} \tanh(h^\star_{\vx} \sqrt{2K}) - \ln\cosh(h^\star_{\vx}\sqrt{2K})\right]
\end{gather}

For arbitrary values of \(K\) Eq.~\eqref{eq:mf-1} has a paramagnetic solution, \(h^\star_{\vx}=0\).
However, the stability of the paramagnetic solution is controlled by the Hessian of the free energy:
\begin{gather}
    M_{\vx,\vy} = \frac{1}{K} f_{\vx,\vy}^{-1} - 2\frac{\delta_{\vx,\vy}}{\cosh^{2}(h^\star_{\vx}\sqrt{2K})}.
\end{gather}
The paramagnetic solution \(h^\star_{\vx} = 0\) is stable as long as the above Hessian remains positive definite: \(M_{\vx,\vy}^{\mathrm{pm}} = \frac{1}{K} f_{\vx,\vy}^{-1} - 2\delta_{\vx,\vy}\).

The critical value of \(K\) for the instability is obtained by looking for the \textit{largest} eigenvalue, \(\lambda_{\mathrm{max}}\), of the linearized equation~\eqref{eq:mf-1} (the operator \(\hat \BG\))
\begin{gather}
    \frac{1}{2K} h^\star _{\vx} = \hat \BG h^\star_{\vy} = \sum_{\vy} f(\vx - \vy) h^\star_{\vy}.
\end{gather}
as \(K=1/(2\lambda_{\mathrm{max}}) \).


\bibliography{general,flatband,frustration,josephson,glass,mbl,josephson1}

\begin{thebibliography}{59}%
\makeatletter
\providecommand \@ifxundefined [1]{%
 \@ifx{#1\undefined}
}%
\providecommand \@ifnum [1]{%
 \ifnum #1\expandafter \@firstoftwo
 \else \expandafter \@secondoftwo
 \fi
}%
\providecommand \@ifx [1]{%
 \ifx #1\expandafter \@firstoftwo
 \else \expandafter \@secondoftwo
 \fi
}%
\providecommand \natexlab [1]{#1}%
\providecommand \enquote  [1]{``#1''}%
\providecommand \bibnamefont  [1]{#1}%
\providecommand \bibfnamefont [1]{#1}%
\providecommand \citenamefont [1]{#1}%
\providecommand \href@noop [0]{\@secondoftwo}%
\providecommand \href [0]{\begingroup \@sanitize@url \@href}%
\providecommand \@href[1]{\@@startlink{#1}\@@href}%
\providecommand \@@href[1]{\endgroup#1\@@endlink}%
\providecommand \@sanitize@url [0]{\catcode `\\12\catcode `\$12\catcode
  `\&12\catcode `\#12\catcode `\^12\catcode `\_12\catcode `\%12\relax}%
\providecommand \@@startlink[1]{}%
\providecommand \@@endlink[0]{}%
\providecommand \url  [0]{\begingroup\@sanitize@url \@url }%
\providecommand \@url [1]{\endgroup\@href {#1}{\urlprefix }}%
\providecommand \urlprefix  [0]{URL }%
\providecommand \Eprint [0]{\href }%
\providecommand \doibase [0]{https://doi.org/}%
\providecommand \selectlanguage [0]{\@gobble}%
\providecommand \bibinfo  [0]{\@secondoftwo}%
\providecommand \bibfield  [0]{\@secondoftwo}%
\providecommand \translation [1]{[#1]}%
\providecommand \BibitemOpen [0]{}%
\providecommand \bibitemStop [0]{}%
\providecommand \bibitemNoStop [0]{.\EOS\space}%
\providecommand \EOS [0]{\spacefactor3000\relax}%
\providecommand \BibitemShut  [1]{\csname bibitem#1\endcsname}%
\let\auto@bib@innerbib\@empty
\bibitem [{\citenamefont {Beloborodov}\ \emph {et~al.}(2007)\citenamefont
  {Beloborodov}, \citenamefont {Lopatin}, \citenamefont {Vinokur},\ and\
  \citenamefont {Efetov}}]{beloborodov2007granular}%
  \BibitemOpen
  \bibfield  {author} {\bibinfo {author} {\bibfnamefont {I.}~\bibnamefont
  {Beloborodov}}, \bibinfo {author} {\bibfnamefont {A.}~\bibnamefont
  {Lopatin}}, \bibinfo {author} {\bibfnamefont {V.}~\bibnamefont {Vinokur}},\
  and\ \bibinfo {author} {\bibfnamefont {K.~B.}\ \bibnamefont {Efetov}},\
  }\bibfield  {title} {\bibinfo {title} {Granular electronic systems},\
  }\href@noop {} {\bibfield  {journal} {\bibinfo  {journal} {Reviews of Modern
  Physics}\ }\textbf {\bibinfo {volume} {79}},\ \bibinfo {pages} {469}
  (\bibinfo {year} {2007})}\BibitemShut {NoStop}%
\bibitem [{\citenamefont {Goldman}(2010)}]{goldman2010superconductor}%
  \BibitemOpen
  \bibfield  {author} {\bibinfo {author} {\bibfnamefont {A.}~\bibnamefont
  {Goldman}},\ }\bibfield  {title} {\bibinfo {title} {Superconductor-insulator
  transitions},\ }\href@noop {} {\bibfield  {journal} {\bibinfo  {journal}
  {International Journal of Modern Physics B}\ }\textbf {\bibinfo {volume}
  {24}},\ \bibinfo {pages} {4081} (\bibinfo {year} {2010})}\BibitemShut
  {NoStop}%
\bibitem [{\citenamefont {Hadley}\ \emph {et~al.}(1988)\citenamefont {Hadley},
  \citenamefont {Beasley},\ and\ \citenamefont {Wiesenfeld}}]{hadley1988phase}%
  \BibitemOpen
  \bibfield  {author} {\bibinfo {author} {\bibfnamefont {P.}~\bibnamefont
  {Hadley}}, \bibinfo {author} {\bibfnamefont {M.}~\bibnamefont {Beasley}},\
  and\ \bibinfo {author} {\bibfnamefont {K.}~\bibnamefont {Wiesenfeld}},\
  }\bibfield  {title} {\bibinfo {title} {Phase locking of josephson-junction
  series arrays},\ }\href@noop {} {\bibfield  {journal} {\bibinfo  {journal}
  {Physical Review B}\ }\textbf {\bibinfo {volume} {38}},\ \bibinfo {pages}
  {8712} (\bibinfo {year} {1988})}\BibitemShut {NoStop}%
\bibitem [{\citenamefont {Rzchowski}(1997)}]{rzchowski1997phase}%
  \BibitemOpen
  \bibfield  {author} {\bibinfo {author} {\bibfnamefont {M.}~\bibnamefont
  {Rzchowski}},\ }\bibfield  {title} {\bibinfo {title} {Phase transitions in a
  kagom\'e lattice of josephson junctions},\ }\href
  {https://doi.org/10.1103/PhysRevB.55.11745} {\bibfield  {journal} {\bibinfo
  {journal} {Phys. Rev. B}\ }\textbf {\bibinfo {volume} {55}},\ \bibinfo
  {pages} {11745} (\bibinfo {year} {1997})}\BibitemShut {NoStop}%
\bibitem [{\citenamefont {van~der Zant}\ \emph {et~al.}(1996)\citenamefont
  {van~der Zant}, \citenamefont {Elion}, \citenamefont {Geerligs},\ and\
  \citenamefont {Mooij}}]{van1996quantum}%
  \BibitemOpen
  \bibfield  {author} {\bibinfo {author} {\bibfnamefont {H.~S.}\ \bibnamefont
  {van~der Zant}}, \bibinfo {author} {\bibfnamefont {W.~J.}\ \bibnamefont
  {Elion}}, \bibinfo {author} {\bibfnamefont {L.~J.}\ \bibnamefont
  {Geerligs}},\ and\ \bibinfo {author} {\bibfnamefont {J.}~\bibnamefont
  {Mooij}},\ }\bibfield  {title} {\bibinfo {title} {Quantum phase transitions
  in two dimensions: Experiments in josephson-junction arrays},\ }\href@noop {}
  {\bibfield  {journal} {\bibinfo  {journal} {Physical Review B}\ }\textbf
  {\bibinfo {volume} {54}},\ \bibinfo {pages} {10081} (\bibinfo {year}
  {1996})}\BibitemShut {NoStop}%
\bibitem [{\citenamefont {Fazio}\ and\ \citenamefont {Van
  Der~Zant}(2001)}]{fazio2001quantum}%
  \BibitemOpen
  \bibfield  {author} {\bibinfo {author} {\bibfnamefont {R.}~\bibnamefont
  {Fazio}}\ and\ \bibinfo {author} {\bibfnamefont {H.}~\bibnamefont {Van
  Der~Zant}},\ }\bibfield  {title} {\bibinfo {title} {Quantum phase transitions
  and vortex dynamics in superconducting networks},\ }\href@noop {} {\bibfield
  {journal} {\bibinfo  {journal} {Physics Reports}\ }\textbf {\bibinfo {volume}
  {355}},\ \bibinfo {pages} {235} (\bibinfo {year} {2001})}\BibitemShut
  {NoStop}%
\bibitem [{\citenamefont {Sondhi}\ \emph {et~al.}(1997)\citenamefont {Sondhi},
  \citenamefont {Girvin}, \citenamefont {Carini},\ and\ \citenamefont
  {Shahar}}]{sondhi1997continuous}%
  \BibitemOpen
  \bibfield  {author} {\bibinfo {author} {\bibfnamefont {S.}~\bibnamefont
  {Sondhi}}, \bibinfo {author} {\bibfnamefont {S.}~\bibnamefont {Girvin}},
  \bibinfo {author} {\bibfnamefont {J.}~\bibnamefont {Carini}},\ and\ \bibinfo
  {author} {\bibfnamefont {D.}~\bibnamefont {Shahar}},\ }\bibfield  {title}
  {\bibinfo {title} {Continuous quantum phase transitions},\ }\href
  {https://doi.org/10.1103/RevModPhys.69.315} {\bibfield  {journal} {\bibinfo
  {journal} {Rev. Mod. Phys.}\ }\textbf {\bibinfo {volume} {69}},\ \bibinfo
  {pages} {315} (\bibinfo {year} {1997})}\BibitemShut {NoStop}%
\bibitem [{\citenamefont {Glazman}\ and\ \citenamefont
  {Larkin}(1997)}]{glazman1997new}%
  \BibitemOpen
  \bibfield  {author} {\bibinfo {author} {\bibfnamefont {L.~I.}\ \bibnamefont
  {Glazman}}\ and\ \bibinfo {author} {\bibfnamefont {A.~I.}\ \bibnamefont
  {Larkin}},\ }\bibfield  {title} {\bibinfo {title} {New quantum phase in a
  one-dimensional josephson array},\ }\href
  {https://doi.org/10.1103/PhysRevLett.79.3736} {\bibfield  {journal} {\bibinfo
   {journal} {Phys. Rev. Lett.}\ }\textbf {\bibinfo {volume} {79}},\ \bibinfo
  {pages} {3736} (\bibinfo {year} {1997})}\BibitemShut {NoStop}%
\bibitem [{\citenamefont {Haviland}\ \emph {et~al.}(2000)\citenamefont
  {Haviland}, \citenamefont {Andersson},\ and\ \citenamefont
  {{\AA}gren}}]{haviland2000superconducting}%
  \BibitemOpen
  \bibfield  {author} {\bibinfo {author} {\bibfnamefont {D.~B.}\ \bibnamefont
  {Haviland}}, \bibinfo {author} {\bibfnamefont {K.}~\bibnamefont
  {Andersson}},\ and\ \bibinfo {author} {\bibfnamefont {P.}~\bibnamefont
  {{\AA}gren}},\ }\bibfield  {title} {\bibinfo {title} {Superconducting and
  insulating behavior in one-dimensional josephson junction arrays},\ }\href
  {https://doi.org/10.1023/A:1004603814529} {\bibfield  {journal} {\bibinfo
  {journal} {Journal of Low Temperature Physics}\ }\textbf {\bibinfo {volume}
  {118}},\ \bibinfo {pages} {733} (\bibinfo {year} {2000})}\BibitemShut
  {NoStop}%
\bibitem [{\citenamefont {Dou\ifmmode~\mbox{\c{c}}\else \c{c}\fi{}ot}\ and\
  \citenamefont {Vidal}(2002)}]{doucot2002pairing}%
  \BibitemOpen
  \bibfield  {author} {\bibinfo {author} {\bibfnamefont {B.}~\bibnamefont
  {Dou\ifmmode~\mbox{\c{c}}\else \c{c}\fi{}ot}}\ and\ \bibinfo {author}
  {\bibfnamefont {J.}~\bibnamefont {Vidal}},\ }\bibfield  {title} {\bibinfo
  {title} {Pairing of cooper pairs in a fully frustrated josephson-junction
  chain},\ }\href {https://doi.org/10.1103/PhysRevLett.88.227005} {\bibfield
  {journal} {\bibinfo  {journal} {Phys. Rev. Lett.}\ }\textbf {\bibinfo
  {volume} {88}},\ \bibinfo {pages} {227005} (\bibinfo {year}
  {2002})}\BibitemShut {NoStop}%
\bibitem [{\citenamefont {Rizzi}\ \emph {et~al.}(2006)\citenamefont {Rizzi},
  \citenamefont {Cataudella},\ and\ \citenamefont {Fazio}}]{rizzi20064}%
  \BibitemOpen
  \bibfield  {author} {\bibinfo {author} {\bibfnamefont {M.}~\bibnamefont
  {Rizzi}}, \bibinfo {author} {\bibfnamefont {V.}~\bibnamefont {Cataudella}},\
  and\ \bibinfo {author} {\bibfnamefont {R.}~\bibnamefont {Fazio}},\ }\bibfield
   {title} {\bibinfo {title} {4 e-condensation in a fully frustrated josephson
  junction diamond chain},\ }\href@noop {} {\bibfield  {journal} {\bibinfo
  {journal} {Physical Review B}\ }\textbf {\bibinfo {volume} {73}},\ \bibinfo
  {pages} {100502} (\bibinfo {year} {2006})}\BibitemShut {NoStop}%
\bibitem [{\citenamefont {Protopopov}\ and\ \citenamefont
  {Feigel'man}(2004)}]{protopopov2004anomalous}%
  \BibitemOpen
  \bibfield  {author} {\bibinfo {author} {\bibfnamefont {I.~V.}\ \bibnamefont
  {Protopopov}}\ and\ \bibinfo {author} {\bibfnamefont {M.~V.}\ \bibnamefont
  {Feigel'man}},\ }\bibfield  {title} {\bibinfo {title} {Anomalous periodicity
  of supercurrent in long frustrated josephson-junction rhombi chains},\ }\href
  {https://doi.org/10.1103/PhysRevB.70.184519} {\bibfield  {journal} {\bibinfo
  {journal} {Phys. Rev. B}\ }\textbf {\bibinfo {volume} {70}},\ \bibinfo
  {pages} {184519} (\bibinfo {year} {2004})}\BibitemShut {NoStop}%
\bibitem [{\citenamefont {Binder}\ \emph {et~al.}(2000)\citenamefont {Binder},
  \citenamefont {Abraimov}, \citenamefont {Ustinov}, \citenamefont {Flach},\
  and\ \citenamefont {Zolotaryuk}}]{binder2000observation}%
  \BibitemOpen
  \bibfield  {author} {\bibinfo {author} {\bibfnamefont {P.}~\bibnamefont
  {Binder}}, \bibinfo {author} {\bibfnamefont {D.}~\bibnamefont {Abraimov}},
  \bibinfo {author} {\bibfnamefont {A.}~\bibnamefont {Ustinov}}, \bibinfo
  {author} {\bibfnamefont {S.}~\bibnamefont {Flach}},\ and\ \bibinfo {author}
  {\bibfnamefont {Y.}~\bibnamefont {Zolotaryuk}},\ }\bibfield  {title}
  {\bibinfo {title} {Observation of breathers in josephson ladders},\ }\href
  {https://doi.org/10.1103/PhysRevLett.84.745} {\bibfield  {journal} {\bibinfo
  {journal} {Phys. Rev. Lett.}\ }\textbf {\bibinfo {volume} {84}},\ \bibinfo
  {pages} {745} (\bibinfo {year} {2000})}\BibitemShut {NoStop}%
\bibitem [{\citenamefont {Trias}\ \emph {et~al.}(2000)\citenamefont {Trias},
  \citenamefont {Mazo},\ and\ \citenamefont {Orlando}}]{trias2000discrete}%
  \BibitemOpen
  \bibfield  {author} {\bibinfo {author} {\bibfnamefont {E.}~\bibnamefont
  {Trias}}, \bibinfo {author} {\bibfnamefont {J.}~\bibnamefont {Mazo}},\ and\
  \bibinfo {author} {\bibfnamefont {T.}~\bibnamefont {Orlando}},\ }\bibfield
  {title} {\bibinfo {title} {Discrete breathers in nonlinear lattices:
  Experimental detection in a josephson array},\ }\href@noop {} {\bibfield
  {journal} {\bibinfo  {journal} {Physical Review Letters}\ }\textbf {\bibinfo
  {volume} {84}},\ \bibinfo {pages} {741} (\bibinfo {year} {2000})}\BibitemShut
  {NoStop}%
\bibitem [{\citenamefont {Ustinov}(1998)}]{ustinov1998solitons}%
  \BibitemOpen
  \bibfield  {author} {\bibinfo {author} {\bibfnamefont {A.}~\bibnamefont
  {Ustinov}},\ }\bibfield  {title} {\bibinfo {title} {Solitons in josephson
  junctions},\ }\href
  {https://doi.org/https://doi.org/10.1016/S0167-2789(98)00131-6} {\bibfield
  {journal} {\bibinfo  {journal} {Phys. D: Nonlin. Phen.}\ }\textbf {\bibinfo
  {volume} {123}},\ \bibinfo {pages} {315 } (\bibinfo {year} {1998})},\
  \bibinfo {note} {annual International Conference of the Center for Nonlinear
  Studies}\BibitemShut {NoStop}%
\bibitem [{\citenamefont {Miroshnichenko}\ \emph {et~al.}(2001)\citenamefont
  {Miroshnichenko}, \citenamefont {Flach}, \citenamefont {Fistul},
  \citenamefont {Zolotaryuk},\ and\ \citenamefont
  {Page}}]{miroshnichenko2001breathers}%
  \BibitemOpen
  \bibfield  {author} {\bibinfo {author} {\bibfnamefont {A.}~\bibnamefont
  {Miroshnichenko}}, \bibinfo {author} {\bibfnamefont {S.}~\bibnamefont
  {Flach}}, \bibinfo {author} {\bibfnamefont {M.}~\bibnamefont {Fistul}},
  \bibinfo {author} {\bibfnamefont {Y.}~\bibnamefont {Zolotaryuk}},\ and\
  \bibinfo {author} {\bibfnamefont {J.}~\bibnamefont {Page}},\ }\bibfield
  {title} {\bibinfo {title} {Breathers in josephson junction ladders:
  Resonances and electromagnetic wave spectroscopy},\ }\href@noop {} {\bibfield
   {journal} {\bibinfo  {journal} {Physical Review E}\ }\textbf {\bibinfo
  {volume} {64}},\ \bibinfo {pages} {066601} (\bibinfo {year}
  {2001})}\BibitemShut {NoStop}%
\bibitem [{\citenamefont {Andreanov}\ and\ \citenamefont
  {Fistul}(2019)}]{andreanov2019resonant}%
  \BibitemOpen
  \bibfield  {author} {\bibinfo {author} {\bibfnamefont {A.}~\bibnamefont
  {Andreanov}}\ and\ \bibinfo {author} {\bibfnamefont {M.~V.}\ \bibnamefont
  {Fistul}},\ }\bibfield  {title} {\bibinfo {title} {Resonant frequencies and
  spatial correlations in frustrated arrays of josephson type nonlinear
  oscillators},\ }\href {https://doi.org/10.1088/1751-8121/ab013d} {\bibfield
  {journal} {\bibinfo  {journal} {J Phys. A: Math. Theor.}\ }\textbf {\bibinfo
  {volume} {52}},\ \bibinfo {pages} {105101} (\bibinfo {year}
  {2019})}\BibitemShut {NoStop}%
\bibitem [{\citenamefont {Matveev}\ \emph {et~al.}(2002)\citenamefont
  {Matveev}, \citenamefont {Larkin},\ and\ \citenamefont
  {Glazman}}]{matveev2002persistent}%
  \BibitemOpen
  \bibfield  {author} {\bibinfo {author} {\bibfnamefont {K.}~\bibnamefont
  {Matveev}}, \bibinfo {author} {\bibfnamefont {A.}~\bibnamefont {Larkin}},\
  and\ \bibinfo {author} {\bibfnamefont {L.}~\bibnamefont {Glazman}},\
  }\bibfield  {title} {\bibinfo {title} {Persistent current in superconducting
  nanorings},\ }\href@noop {} {\bibfield  {journal} {\bibinfo  {journal}
  {Physical review letters}\ }\textbf {\bibinfo {volume} {89}},\ \bibinfo
  {pages} {096802} (\bibinfo {year} {2002})}\BibitemShut {NoStop}%
\bibitem [{\citenamefont {Seidov}\ and\ \citenamefont
  {Fistul}(2021)}]{seidov2021quantum}%
  \BibitemOpen
  \bibfield  {author} {\bibinfo {author} {\bibfnamefont {S.}~\bibnamefont
  {Seidov}}\ and\ \bibinfo {author} {\bibfnamefont {M.}~\bibnamefont
  {Fistul}},\ }\bibfield  {title} {\bibinfo {title} {Quantum dynamics of a
  single fluxon in josephson-junction parallel arrays with large kinetic
  inductances},\ }\href@noop {} {\bibfield  {journal} {\bibinfo  {journal}
  {Physical Review A}\ }\textbf {\bibinfo {volume} {103}},\ \bibinfo {pages}
  {062410} (\bibinfo {year} {2021})}\BibitemShut {NoStop}%
\bibitem [{\citenamefont {Pino}\ \emph {et~al.}(2016)\citenamefont {Pino},
  \citenamefont {Ioffe},\ and\ \citenamefont {Altshuler}}]{pino2016nonergodic}%
  \BibitemOpen
  \bibfield  {author} {\bibinfo {author} {\bibfnamefont {M.}~\bibnamefont
  {Pino}}, \bibinfo {author} {\bibfnamefont {L.~B.}\ \bibnamefont {Ioffe}},\
  and\ \bibinfo {author} {\bibfnamefont {B.~L.}\ \bibnamefont {Altshuler}},\
  }\bibfield  {title} {\bibinfo {title} {Nonergodic metallic and insulating
  phases of josephson junction chains},\ }\href
  {https://doi.org/10.1073/pnas.1520033113} {\bibfield  {journal} {\bibinfo
  {journal} {Proceedings of the National Academy of Sciences}\ }\textbf
  {\bibinfo {volume} {113}},\ \bibinfo {pages} {536} (\bibinfo {year}
  {2016})}\BibitemShut {NoStop}%
\bibitem [{\citenamefont {Schmidt}\ and\ \citenamefont
  {Koch}(2013)}]{schmidt2013circuit}%
  \BibitemOpen
  \bibfield  {author} {\bibinfo {author} {\bibfnamefont {S.}~\bibnamefont
  {Schmidt}}\ and\ \bibinfo {author} {\bibfnamefont {J.}~\bibnamefont {Koch}},\
  }\bibfield  {title} {\bibinfo {title} {Circuit qed lattices: Towards quantum
  simulation with superconducting circuits},\ }\href@noop {} {\bibfield
  {journal} {\bibinfo  {journal} {Annalen der Physik}\ }\textbf {\bibinfo
  {volume} {525}},\ \bibinfo {pages} {395} (\bibinfo {year}
  {2013})}\BibitemShut {NoStop}%
\bibitem [{\citenamefont {Georgescu}\ \emph {et~al.}(2014)\citenamefont
  {Georgescu}, \citenamefont {Ashhab},\ and\ \citenamefont
  {Nori}}]{georgescu2014quantum}%
  \BibitemOpen
  \bibfield  {author} {\bibinfo {author} {\bibfnamefont {I.~M.}\ \bibnamefont
  {Georgescu}}, \bibinfo {author} {\bibfnamefont {S.}~\bibnamefont {Ashhab}},\
  and\ \bibinfo {author} {\bibfnamefont {F.}~\bibnamefont {Nori}},\ }\bibfield
  {title} {\bibinfo {title} {Quantum simulation},\ }\href
  {https://doi.org/10.1103/RevModPhys.86.153} {\bibfield  {journal} {\bibinfo
  {journal} {Rev. Mod. Phys.}\ }\textbf {\bibinfo {volume} {86}},\ \bibinfo
  {pages} {153} (\bibinfo {year} {2014})}\BibitemShut {NoStop}%
\bibitem [{\citenamefont {Acín}\ \emph {et~al.}(2018)\citenamefont {Acín},
  \citenamefont {Bloch}, \citenamefont {Buhrman}, \citenamefont {Calarco},
  \citenamefont {Eichler}, \citenamefont {Eisert}, \citenamefont {Esteve},
  \citenamefont {Gisin}, \citenamefont {Glaser}, \citenamefont {Jelezko},
  \citenamefont {Kuhr}, \citenamefont {Lewenstein}, \citenamefont {Riedel},
  \citenamefont {Schmidt}, \citenamefont {Thew}, \citenamefont {Wallraff},
  \citenamefont {Walmsley},\ and\ \citenamefont {Wilhelm}}]{acin2018quantum}%
  \BibitemOpen
  \bibfield  {author} {\bibinfo {author} {\bibfnamefont {A.}~\bibnamefont
  {Acín}}, \bibinfo {author} {\bibfnamefont {I.}~\bibnamefont {Bloch}},
  \bibinfo {author} {\bibfnamefont {H.}~\bibnamefont {Buhrman}}, \bibinfo
  {author} {\bibfnamefont {T.}~\bibnamefont {Calarco}}, \bibinfo {author}
  {\bibfnamefont {C.}~\bibnamefont {Eichler}}, \bibinfo {author} {\bibfnamefont
  {J.}~\bibnamefont {Eisert}}, \bibinfo {author} {\bibfnamefont
  {D.}~\bibnamefont {Esteve}}, \bibinfo {author} {\bibfnamefont
  {N.}~\bibnamefont {Gisin}}, \bibinfo {author} {\bibfnamefont {S.~J.}\
  \bibnamefont {Glaser}}, \bibinfo {author} {\bibfnamefont {F.}~\bibnamefont
  {Jelezko}}, \bibinfo {author} {\bibfnamefont {S.}~\bibnamefont {Kuhr}},
  \bibinfo {author} {\bibfnamefont {M.}~\bibnamefont {Lewenstein}}, \bibinfo
  {author} {\bibfnamefont {M.~F.}\ \bibnamefont {Riedel}}, \bibinfo {author}
  {\bibfnamefont {P.~O.}\ \bibnamefont {Schmidt}}, \bibinfo {author}
  {\bibfnamefont {R.}~\bibnamefont {Thew}}, \bibinfo {author} {\bibfnamefont
  {A.}~\bibnamefont {Wallraff}}, \bibinfo {author} {\bibfnamefont
  {I.}~\bibnamefont {Walmsley}},\ and\ \bibinfo {author} {\bibfnamefont
  {F.~K.}\ \bibnamefont {Wilhelm}},\ }\bibfield  {title} {\bibinfo {title} {The
  quantum technologies roadmap: a european community view},\ }\href
  {https://doi.org/10.1088/1367-2630/aad1ea} {\bibfield  {journal} {\bibinfo
  {journal} {New Journal of Physics}\ }\textbf {\bibinfo {volume} {20}},\
  \bibinfo {pages} {080201} (\bibinfo {year} {2018})}\BibitemShut {NoStop}%
\bibitem [{\citenamefont {Smith}\ \emph {et~al.}(2019)\citenamefont {Smith},
  \citenamefont {Kim}, \citenamefont {Pollmann},\ and\ \citenamefont
  {Knolle}}]{smith2019simulating}%
  \BibitemOpen
  \bibfield  {author} {\bibinfo {author} {\bibfnamefont {A.}~\bibnamefont
  {Smith}}, \bibinfo {author} {\bibfnamefont {M.~S.}\ \bibnamefont {Kim}},
  \bibinfo {author} {\bibfnamefont {F.}~\bibnamefont {Pollmann}},\ and\
  \bibinfo {author} {\bibfnamefont {J.}~\bibnamefont {Knolle}},\ }\bibfield
  {title} {\bibinfo {title} {Simulating quantum many-body dynamics on a current
  digital quantum computer},\ }\href
  {https://doi.org/10.1038/s41534-019-0217-0} {\bibfield  {journal} {\bibinfo
  {journal} {npj Quantum Information}\ }\textbf {\bibinfo {volume} {5}},\
  \bibinfo {pages} {106} (\bibinfo {year} {2019})}\BibitemShut {NoStop}%
\bibitem [{\citenamefont {Krantz}\ \emph {et~al.}(2019)\citenamefont {Krantz},
  \citenamefont {Kjaergaard}, \citenamefont {Yan}, \citenamefont {Orlando},
  \citenamefont {Gustavsson},\ and\ \citenamefont
  {Oliver}}]{krantz2019quantum}%
  \BibitemOpen
  \bibfield  {author} {\bibinfo {author} {\bibfnamefont {P.}~\bibnamefont
  {Krantz}}, \bibinfo {author} {\bibfnamefont {M.}~\bibnamefont {Kjaergaard}},
  \bibinfo {author} {\bibfnamefont {F.}~\bibnamefont {Yan}}, \bibinfo {author}
  {\bibfnamefont {T.~P.}\ \bibnamefont {Orlando}}, \bibinfo {author}
  {\bibfnamefont {S.}~\bibnamefont {Gustavsson}},\ and\ \bibinfo {author}
  {\bibfnamefont {W.~D.}\ \bibnamefont {Oliver}},\ }\bibfield  {title}
  {\bibinfo {title} {A quantum engineer's guide to superconducting qubits},\
  }\href@noop {} {\bibfield  {journal} {\bibinfo  {journal} {Applied physics
  reviews}\ }\textbf {\bibinfo {volume} {6}} (\bibinfo {year}
  {2019})}\BibitemShut {NoStop}%
\bibitem [{\citenamefont {Kjaergaard}\ \emph {et~al.}(2020)\citenamefont
  {Kjaergaard}, \citenamefont {Schwartz}, \citenamefont {Braum{\"u}ller},
  \citenamefont {Krantz}, \citenamefont {Wang}, \citenamefont {Gustavsson},\
  and\ \citenamefont {Oliver}}]{kjaergaard2020superconducting}%
  \BibitemOpen
  \bibfield  {author} {\bibinfo {author} {\bibfnamefont {M.}~\bibnamefont
  {Kjaergaard}}, \bibinfo {author} {\bibfnamefont {M.~E.}\ \bibnamefont
  {Schwartz}}, \bibinfo {author} {\bibfnamefont {J.}~\bibnamefont
  {Braum{\"u}ller}}, \bibinfo {author} {\bibfnamefont {P.}~\bibnamefont
  {Krantz}}, \bibinfo {author} {\bibfnamefont {J.~I.-J.}\ \bibnamefont {Wang}},
  \bibinfo {author} {\bibfnamefont {S.}~\bibnamefont {Gustavsson}},\ and\
  \bibinfo {author} {\bibfnamefont {W.~D.}\ \bibnamefont {Oliver}},\ }\bibfield
   {title} {\bibinfo {title} {Superconducting qubits: Current state of play},\
  }\href@noop {} {\bibfield  {journal} {\bibinfo  {journal} {Annual Review of
  Condensed Matter Physics}\ }\textbf {\bibinfo {volume} {11}},\ \bibinfo
  {pages} {369} (\bibinfo {year} {2020})}\BibitemShut {NoStop}%
\bibitem [{\citenamefont {Orlando}\ \emph {et~al.}(1999)\citenamefont
  {Orlando}, \citenamefont {Mooij}, \citenamefont {Tian}, \citenamefont {Van
  Der~Wal}, \citenamefont {Levitov}, \citenamefont {Lloyd},\ and\ \citenamefont
  {Mazo}}]{orlando1999superconducting}%
  \BibitemOpen
  \bibfield  {author} {\bibinfo {author} {\bibfnamefont {T.}~\bibnamefont
  {Orlando}}, \bibinfo {author} {\bibfnamefont {J.}~\bibnamefont {Mooij}},
  \bibinfo {author} {\bibfnamefont {L.}~\bibnamefont {Tian}}, \bibinfo {author}
  {\bibfnamefont {C.~H.}\ \bibnamefont {Van Der~Wal}}, \bibinfo {author}
  {\bibfnamefont {L.}~\bibnamefont {Levitov}}, \bibinfo {author} {\bibfnamefont
  {S.}~\bibnamefont {Lloyd}},\ and\ \bibinfo {author} {\bibfnamefont
  {J.}~\bibnamefont {Mazo}},\ }\bibfield  {title} {\bibinfo {title}
  {Superconducting persistent-current qubit},\ }\href@noop {} {\bibfield
  {journal} {\bibinfo  {journal} {Physical Review B}\ }\textbf {\bibinfo
  {volume} {60}},\ \bibinfo {pages} {15398} (\bibinfo {year}
  {1999})}\BibitemShut {NoStop}%
\bibitem [{\citenamefont {Pernack}\ \emph {et~al.}(2024)\citenamefont
  {Pernack}, \citenamefont {Fistul},\ and\ \citenamefont
  {Eremin}}]{pernack2024quantum}%
  \BibitemOpen
  \bibfield  {author} {\bibinfo {author} {\bibfnamefont {B.~J.~P.}\
  \bibnamefont {Pernack}}, \bibinfo {author} {\bibfnamefont {M.~V.}\
  \bibnamefont {Fistul}},\ and\ \bibinfo {author} {\bibfnamefont {I.~M.}\
  \bibnamefont {Eremin}},\ }\bibfield  {title} {\bibinfo {title} {Quantum
  dynamics of frustrated josephson junction arrays embedded in a transmission
  line: An effective $xx$ spin chain with long-range interaction},\ }\href
  {https://doi.org/10.1103/PhysRevB.110.184502} {\bibfield  {journal} {\bibinfo
   {journal} {Phys. Rev. B}\ }\textbf {\bibinfo {volume} {110}},\ \bibinfo
  {pages} {184502} (\bibinfo {year} {2024})}\BibitemShut {NoStop}%
\bibitem [{\citenamefont {Manucharyan}\ \emph {et~al.}(2009)\citenamefont
  {Manucharyan}, \citenamefont {Koch}, \citenamefont {Glazman},\ and\
  \citenamefont {Devoret}}]{manucharyan2009fluxonium}%
  \BibitemOpen
  \bibfield  {author} {\bibinfo {author} {\bibfnamefont {V.~E.}\ \bibnamefont
  {Manucharyan}}, \bibinfo {author} {\bibfnamefont {J.}~\bibnamefont {Koch}},
  \bibinfo {author} {\bibfnamefont {L.~I.}\ \bibnamefont {Glazman}},\ and\
  \bibinfo {author} {\bibfnamefont {M.~H.}\ \bibnamefont {Devoret}},\
  }\bibfield  {title} {\bibinfo {title} {Fluxonium: Single cooper-pair circuit
  free of charge offsets},\ }\href@noop {} {\bibfield  {journal} {\bibinfo
  {journal} {Science}\ }\textbf {\bibinfo {volume} {326}},\ \bibinfo {pages}
  {113} (\bibinfo {year} {2009})}\BibitemShut {NoStop}%
\bibitem [{\citenamefont {Andreanov}\ and\ \citenamefont
  {Fistul}(2020)}]{andreanov2020frustration}%
  \BibitemOpen
  \bibfield  {author} {\bibinfo {author} {\bibfnamefont {A.}~\bibnamefont
  {Andreanov}}\ and\ \bibinfo {author} {\bibfnamefont {M.~V.}\ \bibnamefont
  {Fistul}},\ }\bibfield  {title} {\bibinfo {title} {Frustration-induced highly
  anisotropic magnetic patterns in the classical xy model on the kagome
  lattice},\ }\href {https://doi.org/10.1103/PhysRevB.102.140405} {\bibfield
  {journal} {\bibinfo  {journal} {Phys. Rev. B}\ }\textbf {\bibinfo {volume}
  {102}},\ \bibinfo {pages} {140405} (\bibinfo {year} {2020})}\BibitemShut
  {NoStop}%
\bibitem [{\citenamefont {Neyenhuys}\ \emph {et~al.}(2023)\citenamefont
  {Neyenhuys}, \citenamefont {Fistul},\ and\ \citenamefont
  {Eremin}}]{neyenhuys2023long}%
  \BibitemOpen
  \bibfield  {author} {\bibinfo {author} {\bibfnamefont {O.}~\bibnamefont
  {Neyenhuys}}, \bibinfo {author} {\bibfnamefont {M.~V.}\ \bibnamefont
  {Fistul}},\ and\ \bibinfo {author} {\bibfnamefont {I.~M.}\ \bibnamefont
  {Eremin}},\ }\bibfield  {title} {\bibinfo {title} {Long-range ising spins
  models emerging from frustrated josephson junctions arrays with topological
  constraints},\ }\href@noop {} {\bibfield  {journal} {\bibinfo  {journal}
  {Physical Review B}\ }\textbf {\bibinfo {volume} {108}},\ \bibinfo {pages}
  {165413} (\bibinfo {year} {2023})}\BibitemShut {NoStop}%
\bibitem [{\citenamefont {King}\ \emph {et~al.}(2018)\citenamefont {King},
  \citenamefont {Carrasquilla}, \citenamefont {Raymond}, \citenamefont
  {Ozfidan}, \citenamefont {Andriyash}, \citenamefont {Berkley}, \citenamefont
  {Reis}, \citenamefont {Lanting}, \citenamefont {Harris}, \citenamefont
  {Altomare}, \citenamefont {Boothby}, \citenamefont {Bunyk}, \citenamefont
  {Enderud}, \citenamefont {Fr{\'e}chette}, \citenamefont {Hoskinson},
  \citenamefont {Ladizinsky}, \citenamefont {Oh}, \citenamefont
  {Poulin-Lamarre}, \citenamefont {Rich}, \citenamefont {Sato}, \citenamefont
  {Smirnov}, \citenamefont {Swenson}, \citenamefont {Volkmann}, \citenamefont
  {Whittaker}, \citenamefont {Yao}, \citenamefont {Ladizinsky}, \citenamefont
  {Johnson}, \citenamefont {Hilton},\ and\ \citenamefont
  {Amin}}]{king2018observation}%
  \BibitemOpen
  \bibfield  {author} {\bibinfo {author} {\bibfnamefont {A.~D.}\ \bibnamefont
  {King}}, \bibinfo {author} {\bibfnamefont {J.}~\bibnamefont {Carrasquilla}},
  \bibinfo {author} {\bibfnamefont {J.}~\bibnamefont {Raymond}}, \bibinfo
  {author} {\bibfnamefont {I.}~\bibnamefont {Ozfidan}}, \bibinfo {author}
  {\bibfnamefont {E.}~\bibnamefont {Andriyash}}, \bibinfo {author}
  {\bibfnamefont {A.}~\bibnamefont {Berkley}}, \bibinfo {author} {\bibfnamefont
  {M.}~\bibnamefont {Reis}}, \bibinfo {author} {\bibfnamefont {T.}~\bibnamefont
  {Lanting}}, \bibinfo {author} {\bibfnamefont {R.}~\bibnamefont {Harris}},
  \bibinfo {author} {\bibfnamefont {F.}~\bibnamefont {Altomare}}, \bibinfo
  {author} {\bibfnamefont {K.}~\bibnamefont {Boothby}}, \bibinfo {author}
  {\bibfnamefont {P.~I.}\ \bibnamefont {Bunyk}}, \bibinfo {author}
  {\bibfnamefont {C.}~\bibnamefont {Enderud}}, \bibinfo {author} {\bibfnamefont
  {A.}~\bibnamefont {Fr{\'e}chette}}, \bibinfo {author} {\bibfnamefont
  {E.}~\bibnamefont {Hoskinson}}, \bibinfo {author} {\bibfnamefont
  {N.}~\bibnamefont {Ladizinsky}}, \bibinfo {author} {\bibfnamefont
  {T.}~\bibnamefont {Oh}}, \bibinfo {author} {\bibfnamefont {G.}~\bibnamefont
  {Poulin-Lamarre}}, \bibinfo {author} {\bibfnamefont {C.}~\bibnamefont
  {Rich}}, \bibinfo {author} {\bibfnamefont {Y.}~\bibnamefont {Sato}}, \bibinfo
  {author} {\bibfnamefont {A.~Y.}\ \bibnamefont {Smirnov}}, \bibinfo {author}
  {\bibfnamefont {L.~J.}\ \bibnamefont {Swenson}}, \bibinfo {author}
  {\bibfnamefont {M.~H.}\ \bibnamefont {Volkmann}}, \bibinfo {author}
  {\bibfnamefont {J.}~\bibnamefont {Whittaker}}, \bibinfo {author}
  {\bibfnamefont {J.}~\bibnamefont {Yao}}, \bibinfo {author} {\bibfnamefont
  {E.}~\bibnamefont {Ladizinsky}}, \bibinfo {author} {\bibfnamefont {M.~W.}\
  \bibnamefont {Johnson}}, \bibinfo {author} {\bibfnamefont {J.}~\bibnamefont
  {Hilton}},\ and\ \bibinfo {author} {\bibfnamefont {M.~H.}\ \bibnamefont
  {Amin}},\ }\bibfield  {title} {\bibinfo {title} {Observation of topological
  phenomena in a programmable lattice of 1,800 qubits},\ }\href
  {https://doi.org/10.1038/s41586-018-0410-x} {\bibfield  {journal} {\bibinfo
  {journal} {Nature}\ }\textbf {\bibinfo {volume} {560}},\ \bibinfo {pages}
  {456} (\bibinfo {year} {2018})}\BibitemShut {NoStop}%
\bibitem [{\citenamefont {King}\ \emph {et~al.}(2023)\citenamefont {King},
  \citenamefont {Raymond}, \citenamefont {Lanting}, \citenamefont {Harris},
  \citenamefont {Zucca}, \citenamefont {Altomare}, \citenamefont {Berkley},
  \citenamefont {Boothby}, \citenamefont {Ejtemaee}, \citenamefont {Enderud}
  \emph {et~al.}}]{king2023quantum}%
  \BibitemOpen
  \bibfield  {author} {\bibinfo {author} {\bibfnamefont {A.~D.}\ \bibnamefont
  {King}}, \bibinfo {author} {\bibfnamefont {J.}~\bibnamefont {Raymond}},
  \bibinfo {author} {\bibfnamefont {T.}~\bibnamefont {Lanting}}, \bibinfo
  {author} {\bibfnamefont {R.}~\bibnamefont {Harris}}, \bibinfo {author}
  {\bibfnamefont {A.}~\bibnamefont {Zucca}}, \bibinfo {author} {\bibfnamefont
  {F.}~\bibnamefont {Altomare}}, \bibinfo {author} {\bibfnamefont {A.~J.}\
  \bibnamefont {Berkley}}, \bibinfo {author} {\bibfnamefont {K.}~\bibnamefont
  {Boothby}}, \bibinfo {author} {\bibfnamefont {S.}~\bibnamefont {Ejtemaee}},
  \bibinfo {author} {\bibfnamefont {C.}~\bibnamefont {Enderud}}, \emph
  {et~al.},\ }\bibfield  {title} {\bibinfo {title} {Quantum critical dynamics
  in a 5,000-qubit programmable spin glass},\ }\href@noop {} {\bibfield
  {journal} {\bibinfo  {journal} {Nature}\ ,\ \bibinfo {pages} {1}} (\bibinfo
  {year} {2023})}\BibitemShut {NoStop}%
\bibitem [{\citenamefont {Xu}\ \emph {et~al.}(2018)\citenamefont {Xu},
  \citenamefont {Chen}, \citenamefont {Zeng}, \citenamefont {Zhang},
  \citenamefont {Song}, \citenamefont {Liu}, \citenamefont {Guo}, \citenamefont
  {Zhang}, \citenamefont {Xu}, \citenamefont {Deng}, \citenamefont {Huang},
  \citenamefont {Wang}, \citenamefont {Zhu}, \citenamefont {Zheng},\ and\
  \citenamefont {Fan}}]{xu2018emulating}%
  \BibitemOpen
  \bibfield  {author} {\bibinfo {author} {\bibfnamefont {K.}~\bibnamefont
  {Xu}}, \bibinfo {author} {\bibfnamefont {J.-J.}\ \bibnamefont {Chen}},
  \bibinfo {author} {\bibfnamefont {Y.}~\bibnamefont {Zeng}}, \bibinfo {author}
  {\bibfnamefont {Y.-R.}\ \bibnamefont {Zhang}}, \bibinfo {author}
  {\bibfnamefont {C.}~\bibnamefont {Song}}, \bibinfo {author} {\bibfnamefont
  {W.}~\bibnamefont {Liu}}, \bibinfo {author} {\bibfnamefont {Q.}~\bibnamefont
  {Guo}}, \bibinfo {author} {\bibfnamefont {P.}~\bibnamefont {Zhang}}, \bibinfo
  {author} {\bibfnamefont {D.}~\bibnamefont {Xu}}, \bibinfo {author}
  {\bibfnamefont {H.}~\bibnamefont {Deng}}, \bibinfo {author} {\bibfnamefont
  {K.}~\bibnamefont {Huang}}, \bibinfo {author} {\bibfnamefont
  {H.}~\bibnamefont {Wang}}, \bibinfo {author} {\bibfnamefont {X.}~\bibnamefont
  {Zhu}}, \bibinfo {author} {\bibfnamefont {D.}~\bibnamefont {Zheng}},\ and\
  \bibinfo {author} {\bibfnamefont {H.}~\bibnamefont {Fan}},\ }\bibfield
  {title} {\bibinfo {title} {Emulating many-body localization with a
  superconducting quantum processor},\ }\href
  {https://doi.org/10.1103/PhysRevLett.120.050507} {\bibfield  {journal}
  {\bibinfo  {journal} {Phys. Rev. Lett.}\ }\textbf {\bibinfo {volume} {120}},\
  \bibinfo {pages} {050507} (\bibinfo {year} {2018})}\BibitemShut {NoStop}%
\bibitem [{\citenamefont {Mi}\ \emph {et~al.}(2022)\citenamefont {Mi},
  \citenamefont {Sonner}, \citenamefont {Niu}, \citenamefont {Lee},
  \citenamefont {Foxen}, \citenamefont {Acharya}, \citenamefont {Aleiner},
  \citenamefont {Andersen}, \citenamefont {Arute}, \citenamefont {Arya},
  \citenamefont {Asfaw}, \citenamefont {Atalaya}, \citenamefont {Bardin},
  \citenamefont {Basso}, \citenamefont {Bengtsson}, \citenamefont {Bortoli},
  \citenamefont {Bourassa}, \citenamefont {Brill}, \citenamefont {Broughton},
  \citenamefont {Buckley}, \citenamefont {Buell}, \citenamefont {Burkett},
  \citenamefont {Bushnell}, \citenamefont {Chen}, \citenamefont {Chiaro},
  \citenamefont {Collins}, \citenamefont {Conner}, \citenamefont {Courtney},
  \citenamefont {Crook}, \citenamefont {Debroy}, \citenamefont {Demura},
  \citenamefont {Dunsworth}, \citenamefont {Eppens}, \citenamefont {Erickson},
  \citenamefont {Faoro}, \citenamefont {Farhi}, \citenamefont {Fatemi},
  \citenamefont {Flores}, \citenamefont {Forati}, \citenamefont {Fowler},
  \citenamefont {Giang}, \citenamefont {Gidney}, \citenamefont {Gilboa},
  \citenamefont {Giustina}, \citenamefont {Dau}, \citenamefont {Gross},
  \citenamefont {Habegger}, \citenamefont {Harrigan}, \citenamefont {Hoffmann},
  \citenamefont {Hong}, \citenamefont {Huang}, \citenamefont {Huff},
  \citenamefont {Huggins}, \citenamefont {Ioffe}, \citenamefont {Isakov},
  \citenamefont {Iveland}, \citenamefont {Jeffrey}, \citenamefont {Jiang},
  \citenamefont {Jones}, \citenamefont {Kafri}, \citenamefont {Kechedzhi},
  \citenamefont {Khattar}, \citenamefont {Kim}, \citenamefont {Kitaev},
  \citenamefont {Klimov}, \citenamefont {Klots}, \citenamefont {Korotkov},
  \citenamefont {Kostritsa}, \citenamefont {Kreikebaum}, \citenamefont
  {Landhuis}, \citenamefont {Laptev}, \citenamefont {Lau}, \citenamefont {Lee},
  \citenamefont {Laws}, \citenamefont {Liu}, \citenamefont {Locharla},
  \citenamefont {Martin}, \citenamefont {McClean}, \citenamefont {McEwen},
  \citenamefont {Costa}, \citenamefont {Miao}, \citenamefont {Mohseni},
  \citenamefont {Montazeri}, \citenamefont {Morvan}, \citenamefont {Mount},
  \citenamefont {Mruczkiewicz}, \citenamefont {Naaman}, \citenamefont {Neeley},
  \citenamefont {Neill}, \citenamefont {Newman}, \citenamefont {O’Brien},
  \citenamefont {Opremcak}, \citenamefont {Petukhov}, \citenamefont {Potter},
  \citenamefont {Quintana}, \citenamefont {Rubin}, \citenamefont {Saei},
  \citenamefont {Sank}, \citenamefont {Sankaragomathi}, \citenamefont
  {Satzinger}, \citenamefont {Schuster}, \citenamefont {Shearn}, \citenamefont
  {Shvarts}, \citenamefont {Strain}, \citenamefont {Su}, \citenamefont
  {Szalay}, \citenamefont {Vidal}, \citenamefont {Villalonga}, \citenamefont
  {Vollgraff-Heidweiller}, \citenamefont {White}, \citenamefont {Yao},
  \citenamefont {Yeh}, \citenamefont {Yoo}, \citenamefont {Zalcman},
  \citenamefont {Zhang}, \citenamefont {Zhu}, \citenamefont {Neven},
  \citenamefont {Bacon}, \citenamefont {Hilton}, \citenamefont {Lucero},
  \citenamefont {Babbush}, \citenamefont {Boixo}, \citenamefont {Megrant},
  \citenamefont {Chen}, \citenamefont {Kelly}, \citenamefont {Smelyanskiy},
  \citenamefont {Abanin},\ and\ \citenamefont {Roushan}}]{mi2022noise}%
  \BibitemOpen
  \bibfield  {author} {\bibinfo {author} {\bibfnamefont {X.}~\bibnamefont
  {Mi}}, \bibinfo {author} {\bibfnamefont {M.}~\bibnamefont {Sonner}}, \bibinfo
  {author} {\bibfnamefont {M.~Y.}\ \bibnamefont {Niu}}, \bibinfo {author}
  {\bibfnamefont {K.~W.}\ \bibnamefont {Lee}}, \bibinfo {author} {\bibfnamefont
  {B.}~\bibnamefont {Foxen}}, \bibinfo {author} {\bibfnamefont
  {R.}~\bibnamefont {Acharya}}, \bibinfo {author} {\bibfnamefont
  {I.}~\bibnamefont {Aleiner}}, \bibinfo {author} {\bibfnamefont {T.~I.}\
  \bibnamefont {Andersen}}, \bibinfo {author} {\bibfnamefont {F.}~\bibnamefont
  {Arute}}, \bibinfo {author} {\bibfnamefont {K.}~\bibnamefont {Arya}},
  \bibinfo {author} {\bibfnamefont {A.}~\bibnamefont {Asfaw}}, \bibinfo
  {author} {\bibfnamefont {J.}~\bibnamefont {Atalaya}}, \bibinfo {author}
  {\bibfnamefont {J.~C.}\ \bibnamefont {Bardin}}, \bibinfo {author}
  {\bibfnamefont {J.}~\bibnamefont {Basso}}, \bibinfo {author} {\bibfnamefont
  {A.}~\bibnamefont {Bengtsson}}, \bibinfo {author} {\bibfnamefont
  {G.}~\bibnamefont {Bortoli}}, \bibinfo {author} {\bibfnamefont
  {A.}~\bibnamefont {Bourassa}}, \bibinfo {author} {\bibfnamefont
  {L.}~\bibnamefont {Brill}}, \bibinfo {author} {\bibfnamefont
  {M.}~\bibnamefont {Broughton}}, \bibinfo {author} {\bibfnamefont {B.~B.}\
  \bibnamefont {Buckley}}, \bibinfo {author} {\bibfnamefont {D.~A.}\
  \bibnamefont {Buell}}, \bibinfo {author} {\bibfnamefont {B.}~\bibnamefont
  {Burkett}}, \bibinfo {author} {\bibfnamefont {N.}~\bibnamefont {Bushnell}},
  \bibinfo {author} {\bibfnamefont {Z.}~\bibnamefont {Chen}}, \bibinfo {author}
  {\bibfnamefont {B.}~\bibnamefont {Chiaro}}, \bibinfo {author} {\bibfnamefont
  {R.}~\bibnamefont {Collins}}, \bibinfo {author} {\bibfnamefont
  {P.}~\bibnamefont {Conner}}, \bibinfo {author} {\bibfnamefont
  {W.}~\bibnamefont {Courtney}}, \bibinfo {author} {\bibfnamefont {A.~L.}\
  \bibnamefont {Crook}}, \bibinfo {author} {\bibfnamefont {D.~M.}\ \bibnamefont
  {Debroy}}, \bibinfo {author} {\bibfnamefont {S.}~\bibnamefont {Demura}},
  \bibinfo {author} {\bibfnamefont {A.}~\bibnamefont {Dunsworth}}, \bibinfo
  {author} {\bibfnamefont {D.}~\bibnamefont {Eppens}}, \bibinfo {author}
  {\bibfnamefont {C.}~\bibnamefont {Erickson}}, \bibinfo {author}
  {\bibfnamefont {L.}~\bibnamefont {Faoro}}, \bibinfo {author} {\bibfnamefont
  {E.}~\bibnamefont {Farhi}}, \bibinfo {author} {\bibfnamefont
  {R.}~\bibnamefont {Fatemi}}, \bibinfo {author} {\bibfnamefont
  {L.}~\bibnamefont {Flores}}, \bibinfo {author} {\bibfnamefont
  {E.}~\bibnamefont {Forati}}, \bibinfo {author} {\bibfnamefont {A.~G.}\
  \bibnamefont {Fowler}}, \bibinfo {author} {\bibfnamefont {W.}~\bibnamefont
  {Giang}}, \bibinfo {author} {\bibfnamefont {C.}~\bibnamefont {Gidney}},
  \bibinfo {author} {\bibfnamefont {D.}~\bibnamefont {Gilboa}}, \bibinfo
  {author} {\bibfnamefont {M.}~\bibnamefont {Giustina}}, \bibinfo {author}
  {\bibfnamefont {A.~G.}\ \bibnamefont {Dau}}, \bibinfo {author} {\bibfnamefont
  {J.~A.}\ \bibnamefont {Gross}}, \bibinfo {author} {\bibfnamefont
  {S.}~\bibnamefont {Habegger}}, \bibinfo {author} {\bibfnamefont {M.~P.}\
  \bibnamefont {Harrigan}}, \bibinfo {author} {\bibfnamefont {M.}~\bibnamefont
  {Hoffmann}}, \bibinfo {author} {\bibfnamefont {S.}~\bibnamefont {Hong}},
  \bibinfo {author} {\bibfnamefont {T.}~\bibnamefont {Huang}}, \bibinfo
  {author} {\bibfnamefont {A.}~\bibnamefont {Huff}}, \bibinfo {author}
  {\bibfnamefont {W.~J.}\ \bibnamefont {Huggins}}, \bibinfo {author}
  {\bibfnamefont {L.~B.}\ \bibnamefont {Ioffe}}, \bibinfo {author}
  {\bibfnamefont {S.~V.}\ \bibnamefont {Isakov}}, \bibinfo {author}
  {\bibfnamefont {J.}~\bibnamefont {Iveland}}, \bibinfo {author} {\bibfnamefont
  {E.}~\bibnamefont {Jeffrey}}, \bibinfo {author} {\bibfnamefont
  {Z.}~\bibnamefont {Jiang}}, \bibinfo {author} {\bibfnamefont
  {C.}~\bibnamefont {Jones}}, \bibinfo {author} {\bibfnamefont
  {D.}~\bibnamefont {Kafri}}, \bibinfo {author} {\bibfnamefont
  {K.}~\bibnamefont {Kechedzhi}}, \bibinfo {author} {\bibfnamefont
  {T.}~\bibnamefont {Khattar}}, \bibinfo {author} {\bibfnamefont
  {S.}~\bibnamefont {Kim}}, \bibinfo {author} {\bibfnamefont {A.~Y.}\
  \bibnamefont {Kitaev}}, \bibinfo {author} {\bibfnamefont {P.~V.}\
  \bibnamefont {Klimov}}, \bibinfo {author} {\bibfnamefont {A.~R.}\
  \bibnamefont {Klots}}, \bibinfo {author} {\bibfnamefont {A.~N.}\ \bibnamefont
  {Korotkov}}, \bibinfo {author} {\bibfnamefont {F.}~\bibnamefont {Kostritsa}},
  \bibinfo {author} {\bibfnamefont {J.~M.}\ \bibnamefont {Kreikebaum}},
  \bibinfo {author} {\bibfnamefont {D.}~\bibnamefont {Landhuis}}, \bibinfo
  {author} {\bibfnamefont {P.}~\bibnamefont {Laptev}}, \bibinfo {author}
  {\bibfnamefont {K.-M.}\ \bibnamefont {Lau}}, \bibinfo {author} {\bibfnamefont
  {J.}~\bibnamefont {Lee}}, \bibinfo {author} {\bibfnamefont {L.}~\bibnamefont
  {Laws}}, \bibinfo {author} {\bibfnamefont {W.}~\bibnamefont {Liu}}, \bibinfo
  {author} {\bibfnamefont {A.}~\bibnamefont {Locharla}}, \bibinfo {author}
  {\bibfnamefont {O.}~\bibnamefont {Martin}}, \bibinfo {author} {\bibfnamefont
  {J.~R.}\ \bibnamefont {McClean}}, \bibinfo {author} {\bibfnamefont
  {M.}~\bibnamefont {McEwen}}, \bibinfo {author} {\bibfnamefont {B.~M.}\
  \bibnamefont {Costa}}, \bibinfo {author} {\bibfnamefont {K.~C.}\ \bibnamefont
  {Miao}}, \bibinfo {author} {\bibfnamefont {M.}~\bibnamefont {Mohseni}},
  \bibinfo {author} {\bibfnamefont {S.}~\bibnamefont {Montazeri}}, \bibinfo
  {author} {\bibfnamefont {A.}~\bibnamefont {Morvan}}, \bibinfo {author}
  {\bibfnamefont {E.}~\bibnamefont {Mount}}, \bibinfo {author} {\bibfnamefont
  {W.}~\bibnamefont {Mruczkiewicz}}, \bibinfo {author} {\bibfnamefont
  {O.}~\bibnamefont {Naaman}}, \bibinfo {author} {\bibfnamefont
  {M.}~\bibnamefont {Neeley}}, \bibinfo {author} {\bibfnamefont
  {C.}~\bibnamefont {Neill}}, \bibinfo {author} {\bibfnamefont
  {M.}~\bibnamefont {Newman}}, \bibinfo {author} {\bibfnamefont {T.~E.}\
  \bibnamefont {O’Brien}}, \bibinfo {author} {\bibfnamefont {A.}~\bibnamefont
  {Opremcak}}, \bibinfo {author} {\bibfnamefont {A.}~\bibnamefont {Petukhov}},
  \bibinfo {author} {\bibfnamefont {R.}~\bibnamefont {Potter}}, \bibinfo
  {author} {\bibfnamefont {C.}~\bibnamefont {Quintana}}, \bibinfo {author}
  {\bibfnamefont {N.~C.}\ \bibnamefont {Rubin}}, \bibinfo {author}
  {\bibfnamefont {N.}~\bibnamefont {Saei}}, \bibinfo {author} {\bibfnamefont
  {D.}~\bibnamefont {Sank}}, \bibinfo {author} {\bibfnamefont {K.}~\bibnamefont
  {Sankaragomathi}}, \bibinfo {author} {\bibfnamefont {K.~J.}\ \bibnamefont
  {Satzinger}}, \bibinfo {author} {\bibfnamefont {C.}~\bibnamefont {Schuster}},
  \bibinfo {author} {\bibfnamefont {M.~J.}\ \bibnamefont {Shearn}}, \bibinfo
  {author} {\bibfnamefont {V.}~\bibnamefont {Shvarts}}, \bibinfo {author}
  {\bibfnamefont {D.}~\bibnamefont {Strain}}, \bibinfo {author} {\bibfnamefont
  {Y.}~\bibnamefont {Su}}, \bibinfo {author} {\bibfnamefont {M.}~\bibnamefont
  {Szalay}}, \bibinfo {author} {\bibfnamefont {G.}~\bibnamefont {Vidal}},
  \bibinfo {author} {\bibfnamefont {B.}~\bibnamefont {Villalonga}}, \bibinfo
  {author} {\bibfnamefont {C.}~\bibnamefont {Vollgraff-Heidweiller}}, \bibinfo
  {author} {\bibfnamefont {T.}~\bibnamefont {White}}, \bibinfo {author}
  {\bibfnamefont {Z.}~\bibnamefont {Yao}}, \bibinfo {author} {\bibfnamefont
  {P.}~\bibnamefont {Yeh}}, \bibinfo {author} {\bibfnamefont {J.}~\bibnamefont
  {Yoo}}, \bibinfo {author} {\bibfnamefont {A.}~\bibnamefont {Zalcman}},
  \bibinfo {author} {\bibfnamefont {Y.}~\bibnamefont {Zhang}}, \bibinfo
  {author} {\bibfnamefont {N.}~\bibnamefont {Zhu}}, \bibinfo {author}
  {\bibfnamefont {H.}~\bibnamefont {Neven}}, \bibinfo {author} {\bibfnamefont
  {D.}~\bibnamefont {Bacon}}, \bibinfo {author} {\bibfnamefont
  {J.}~\bibnamefont {Hilton}}, \bibinfo {author} {\bibfnamefont
  {E.}~\bibnamefont {Lucero}}, \bibinfo {author} {\bibfnamefont
  {R.}~\bibnamefont {Babbush}}, \bibinfo {author} {\bibfnamefont
  {S.}~\bibnamefont {Boixo}}, \bibinfo {author} {\bibfnamefont
  {A.}~\bibnamefont {Megrant}}, \bibinfo {author} {\bibfnamefont
  {Y.}~\bibnamefont {Chen}}, \bibinfo {author} {\bibfnamefont {J.}~\bibnamefont
  {Kelly}}, \bibinfo {author} {\bibfnamefont {V.}~\bibnamefont {Smelyanskiy}},
  \bibinfo {author} {\bibfnamefont {D.~A.}\ \bibnamefont {Abanin}},\ and\
  \bibinfo {author} {\bibfnamefont {P.}~\bibnamefont {Roushan}},\ }\bibfield
  {title} {\bibinfo {title} {Noise-resilient edge modes on a chain of
  superconducting qubits},\ }\href {https://doi.org/10.1126/science.abq5769}
  {\bibfield  {journal} {\bibinfo  {journal} {Science}\ }\textbf {\bibinfo
  {volume} {378}},\ \bibinfo {pages} {785} (\bibinfo {year}
  {2022})}\BibitemShut {NoStop}%
\bibitem [{\citenamefont {Heras}\ \emph {et~al.}(2014)\citenamefont {Heras},
  \citenamefont {Mezzacapo}, \citenamefont {Lamata}, \citenamefont {Filipp},
  \citenamefont {Wallraff},\ and\ \citenamefont {Solano}}]{las2014digital}%
  \BibitemOpen
  \bibfield  {author} {\bibinfo {author} {\bibfnamefont {U.~L.}\ \bibnamefont
  {Heras}}, \bibinfo {author} {\bibfnamefont {A.}~\bibnamefont {Mezzacapo}},
  \bibinfo {author} {\bibfnamefont {L.}~\bibnamefont {Lamata}}, \bibinfo
  {author} {\bibfnamefont {S.}~\bibnamefont {Filipp}}, \bibinfo {author}
  {\bibfnamefont {A.}~\bibnamefont {Wallraff}},\ and\ \bibinfo {author}
  {\bibfnamefont {E.}~\bibnamefont {Solano}},\ }\bibfield  {title} {\bibinfo
  {title} {Digital quantum simulation of spin systems in superconducting
  circuits},\ }\href {https://doi.org/10.1103/PhysRevLett.112.200501}
  {\bibfield  {journal} {\bibinfo  {journal} {Phys. Rev. Lett.}\ }\textbf
  {\bibinfo {volume} {112}},\ \bibinfo {pages} {200501} (\bibinfo {year}
  {2014})}\BibitemShut {NoStop}%
\bibitem [{\citenamefont {Park}\ and\ \citenamefont
  {Lee}(2022)}]{park2022frustrated}%
  \BibitemOpen
  \bibfield  {author} {\bibinfo {author} {\bibfnamefont {H.}~\bibnamefont
  {Park}}\ and\ \bibinfo {author} {\bibfnamefont {H.}~\bibnamefont {Lee}},\
  }\bibfield  {title} {\bibinfo {title} {Frustrated ising model on d-wave
  quantum annealing machine},\ }\href {https://doi.org/10.7566/JPSJ.91.074001}
  {\bibfield  {journal} {\bibinfo  {journal} {Journal of the Physical Society
  of Japan}\ }\textbf {\bibinfo {volume} {91}},\ \bibinfo {pages} {074001}
  (\bibinfo {year} {2022})}\BibitemShut {NoStop}%
\bibitem [{\citenamefont {Anderson}(1978)}]{anderson1978concept}%
  \BibitemOpen
  \bibfield  {author} {\bibinfo {author} {\bibfnamefont {P.}~\bibnamefont
  {Anderson}},\ }\bibfield  {title} {\bibinfo {title} {The concept of
  frustration in spin glasses},\ }\href
  {https://doi.org/https://doi.org/10.1016/0022-5088(78)90040-1} {\bibfield
  {journal} {\bibinfo  {journal} {Journal of the Less Common Metals}\ }\textbf
  {\bibinfo {volume} {62}},\ \bibinfo {pages} {291} (\bibinfo {year}
  {1978})}\BibitemShut {NoStop}%
\bibitem [{\citenamefont {Nisoli}\ \emph {et~al.}(2013)\citenamefont {Nisoli},
  \citenamefont {Moessner},\ and\ \citenamefont
  {Schiffer}}]{nisoli2013colloquium}%
  \BibitemOpen
  \bibfield  {author} {\bibinfo {author} {\bibfnamefont {C.}~\bibnamefont
  {Nisoli}}, \bibinfo {author} {\bibfnamefont {R.}~\bibnamefont {Moessner}},\
  and\ \bibinfo {author} {\bibfnamefont {P.}~\bibnamefont {Schiffer}},\
  }\bibfield  {title} {\bibinfo {title} {Colloquium: Artificial spin ice:
  Designing and imaging magnetic frustration},\ }\href
  {https://doi.org/10.1103/RevModPhys.85.1473} {\bibfield  {journal} {\bibinfo
  {journal} {Rev. Mod. Phys.}\ }\textbf {\bibinfo {volume} {85}},\ \bibinfo
  {pages} {1473} (\bibinfo {year} {2013})}\BibitemShut {NoStop}%
\bibitem [{\citenamefont {Moessner}\ and\ \citenamefont
  {Ramirez}(2006)}]{moessner2006geometrical}%
  \BibitemOpen
  \bibfield  {author} {\bibinfo {author} {\bibfnamefont {R.}~\bibnamefont
  {Moessner}}\ and\ \bibinfo {author} {\bibfnamefont {A.~P.}\ \bibnamefont
  {Ramirez}},\ }\bibfield  {title} {\bibinfo {title} {Geometrical
  frustration},\ }\href@noop {} {\bibfield  {journal} {\bibinfo  {journal}
  {Physics Today}\ }\textbf {\bibinfo {volume} {59}},\ \bibinfo {pages} {24}
  (\bibinfo {year} {2006})}\BibitemShut {NoStop}%
\bibitem [{\citenamefont {Schr\"oder}\ \emph {et~al.}(2005)\citenamefont
  {Schr\"oder}, \citenamefont {Nojiri}, \citenamefont {Schnack}, \citenamefont
  {Hage}, \citenamefont {Luban},\ and\ \citenamefont
  {K\"ogerler}}]{schroeder2005competing}%
  \BibitemOpen
  \bibfield  {author} {\bibinfo {author} {\bibfnamefont {C.}~\bibnamefont
  {Schr\"oder}}, \bibinfo {author} {\bibfnamefont {H.}~\bibnamefont {Nojiri}},
  \bibinfo {author} {\bibfnamefont {J.}~\bibnamefont {Schnack}}, \bibinfo
  {author} {\bibfnamefont {P.}~\bibnamefont {Hage}}, \bibinfo {author}
  {\bibfnamefont {M.}~\bibnamefont {Luban}},\ and\ \bibinfo {author}
  {\bibfnamefont {P.}~\bibnamefont {K\"ogerler}},\ }\bibfield  {title}
  {\bibinfo {title} {Competing spin phases in geometrically frustrated magnetic
  molecules},\ }\href {https://doi.org/10.1103/PhysRevLett.94.017205}
  {\bibfield  {journal} {\bibinfo  {journal} {Phys. Rev. Lett.}\ }\textbf
  {\bibinfo {volume} {94}},\ \bibinfo {pages} {017205} (\bibinfo {year}
  {2005})}\BibitemShut {NoStop}%
\bibitem [{\citenamefont {Balents}(2010)}]{balents2010spin}%
  \BibitemOpen
  \bibfield  {author} {\bibinfo {author} {\bibfnamefont {L.}~\bibnamefont
  {Balents}},\ }\bibfield  {title} {\bibinfo {title} {Spin liquids in
  frustrated magnets},\ }\href {https://doi.org/10.1038/nature08917} {\bibfield
   {journal} {\bibinfo  {journal} {Nature}\ }\textbf {\bibinfo {volume}
  {464}},\ \bibinfo {pages} {199} (\bibinfo {year} {2010})}\BibitemShut
  {NoStop}%
\bibitem [{\citenamefont {Baniodeh}\ \emph {et~al.}(2018)\citenamefont
  {Baniodeh}, \citenamefont {Magnani}, \citenamefont {Lan}, \citenamefont
  {Buth}, \citenamefont {Anson}, \citenamefont {Richter}, \citenamefont
  {Affronte}, \citenamefont {Schnack},\ and\ \citenamefont
  {Powell}}]{baniodeh2018high}%
  \BibitemOpen
  \bibfield  {author} {\bibinfo {author} {\bibfnamefont {A.}~\bibnamefont
  {Baniodeh}}, \bibinfo {author} {\bibfnamefont {N.}~\bibnamefont {Magnani}},
  \bibinfo {author} {\bibfnamefont {Y.}~\bibnamefont {Lan}}, \bibinfo {author}
  {\bibfnamefont {G.}~\bibnamefont {Buth}}, \bibinfo {author} {\bibfnamefont
  {C.~E.}\ \bibnamefont {Anson}}, \bibinfo {author} {\bibfnamefont
  {J.}~\bibnamefont {Richter}}, \bibinfo {author} {\bibfnamefont
  {M.}~\bibnamefont {Affronte}}, \bibinfo {author} {\bibfnamefont
  {J.}~\bibnamefont {Schnack}},\ and\ \bibinfo {author} {\bibfnamefont {A.~K.}\
  \bibnamefont {Powell}},\ }\bibfield  {title} {\bibinfo {title} {High spin
  cycles: topping the spin record for a single molecule verging on quantum
  criticality},\ }\href@noop {} {\bibfield  {journal} {\bibinfo  {journal} {npj
  Quantum Materials}\ }\textbf {\bibinfo {volume} {3}},\ \bibinfo {pages} {10}
  (\bibinfo {year} {2018})}\BibitemShut {NoStop}%
\bibitem [{\citenamefont {Han}\ \emph {et~al.}(2012)\citenamefont {Han},
  \citenamefont {Helton}, \citenamefont {Chu}, \citenamefont {Nocera},
  \citenamefont {Rodriguez-Rivera}, \citenamefont {Broholm},\ and\
  \citenamefont {Lee}}]{han2012fractionalized}%
  \BibitemOpen
  \bibfield  {author} {\bibinfo {author} {\bibfnamefont {T.-H.}\ \bibnamefont
  {Han}}, \bibinfo {author} {\bibfnamefont {J.~S.}\ \bibnamefont {Helton}},
  \bibinfo {author} {\bibfnamefont {S.}~\bibnamefont {Chu}}, \bibinfo {author}
  {\bibfnamefont {D.~G.}\ \bibnamefont {Nocera}}, \bibinfo {author}
  {\bibfnamefont {J.~A.}\ \bibnamefont {Rodriguez-Rivera}}, \bibinfo {author}
  {\bibfnamefont {C.}~\bibnamefont {Broholm}},\ and\ \bibinfo {author}
  {\bibfnamefont {Y.~S.}\ \bibnamefont {Lee}},\ }\bibfield  {title} {\bibinfo
  {title} {Fractionalized excitations in the spin-liquid state of a
  kagome-lattice antiferromagnet},\ }\href
  {https://doi.org/10.1038/nature11659} {\bibfield  {journal} {\bibinfo
  {journal} {Nature}\ }\textbf {\bibinfo {volume} {492}},\ \bibinfo {pages}
  {406} (\bibinfo {year} {2012})}\BibitemShut {NoStop}%
\bibitem [{\citenamefont {Mahmoudian}\ \emph {et~al.}(2015)\citenamefont
  {Mahmoudian}, \citenamefont {Rademaker}, \citenamefont {Ralko}, \citenamefont
  {Fratini},\ and\ \citenamefont {Dobrosavljevi\ifmmode~\acute{c}\else
  \'{c}\fi{}}}]{mahmoudian2015glassy}%
  \BibitemOpen
  \bibfield  {author} {\bibinfo {author} {\bibfnamefont {S.}~\bibnamefont
  {Mahmoudian}}, \bibinfo {author} {\bibfnamefont {L.}~\bibnamefont
  {Rademaker}}, \bibinfo {author} {\bibfnamefont {A.}~\bibnamefont {Ralko}},
  \bibinfo {author} {\bibfnamefont {S.}~\bibnamefont {Fratini}},\ and\ \bibinfo
  {author} {\bibfnamefont {V.}~\bibnamefont
  {Dobrosavljevi\ifmmode~\acute{c}\else \'{c}\fi{}}},\ }\bibfield  {title}
  {\bibinfo {title} {Glassy dynamics in geometrically frustrated coulomb
  liquids without disorder},\ }\href
  {https://doi.org/10.1103/PhysRevLett.115.025701} {\bibfield  {journal}
  {\bibinfo  {journal} {Phys. Rev. Lett.}\ }\textbf {\bibinfo {volume} {115}},\
  \bibinfo {pages} {025701} (\bibinfo {year} {2015})}\BibitemShut {NoStop}%
\bibitem [{\citenamefont {Caputo}\ \emph {et~al.}(2001)\citenamefont {Caputo},
  \citenamefont {Fistul},\ and\ \citenamefont
  {Ustinov}}]{caputo2001resonances}%
  \BibitemOpen
  \bibfield  {author} {\bibinfo {author} {\bibfnamefont {P.}~\bibnamefont
  {Caputo}}, \bibinfo {author} {\bibfnamefont {M.}~\bibnamefont {Fistul}},\
  and\ \bibinfo {author} {\bibfnamefont {A.}~\bibnamefont {Ustinov}},\
  }\bibfield  {title} {\bibinfo {title} {Resonances in one and two rows of
  triangular josephson junction cells},\ }\href
  {https://doi.org/10.1103/PhysRevB.63.214510} {\bibfield  {journal} {\bibinfo
  {journal} {Phys. Rev. B}\ }\textbf {\bibinfo {volume} {63}},\ \bibinfo
  {pages} {214510} (\bibinfo {year} {2001})}\BibitemShut {NoStop}%
\bibitem [{\citenamefont {Valdez-Balderas}\ and\ \citenamefont
  {Stroud}(2005)}]{valdez2005superconductivity}%
  \BibitemOpen
  \bibfield  {author} {\bibinfo {author} {\bibfnamefont {D.}~\bibnamefont
  {Valdez-Balderas}}\ and\ \bibinfo {author} {\bibfnamefont {D.}~\bibnamefont
  {Stroud}},\ }\bibfield  {title} {\bibinfo {title} {{Superconductivity versus
  phase separation, stripes, and checkerboard ordering: A two-dimensional Monte
  Carlo study}},\ }\href {https://doi.org/10.1103/PhysRevB.72.214501}
  {\bibfield  {journal} {\bibinfo  {journal} {Phys. Rev. B}\ }\textbf {\bibinfo
  {volume} {72}},\ \bibinfo {pages} {214501} (\bibinfo {year}
  {2005})}\BibitemShut {NoStop}%
\bibitem [{\citenamefont {Pop}\ \emph {et~al.}(2008)\citenamefont {Pop},
  \citenamefont {Hasselbach}, \citenamefont {Buisson}, \citenamefont
  {Guichard}, \citenamefont {Pannetier},\ and\ \citenamefont
  {Protopopov}}]{pop2008measurement}%
  \BibitemOpen
  \bibfield  {author} {\bibinfo {author} {\bibfnamefont {I.}~\bibnamefont
  {Pop}}, \bibinfo {author} {\bibfnamefont {K.}~\bibnamefont {Hasselbach}},
  \bibinfo {author} {\bibfnamefont {O.}~\bibnamefont {Buisson}}, \bibinfo
  {author} {\bibfnamefont {W.}~\bibnamefont {Guichard}}, \bibinfo {author}
  {\bibfnamefont {B.}~\bibnamefont {Pannetier}},\ and\ \bibinfo {author}
  {\bibfnamefont {I.}~\bibnamefont {Protopopov}},\ }\bibfield  {title}
  {\bibinfo {title} {Measurement of the current-phase relation in josephson
  junction rhombi chains},\ }\href {https://doi.org/10.1103/PhysRevB.78.104504}
  {\bibfield  {journal} {\bibinfo  {journal} {Phys. Rev. B}\ }\textbf {\bibinfo
  {volume} {78}},\ \bibinfo {pages} {104504} (\bibinfo {year}
  {2008})}\BibitemShut {NoStop}%
\bibitem [{\citenamefont {Feofanov}\ \emph {et~al.}(2010)\citenamefont
  {Feofanov}, \citenamefont {Oboznov}, \citenamefont {Bol'ginov}, \citenamefont
  {Lisenfeld}, \citenamefont {Poletto}, \citenamefont {Ryazanov}, \citenamefont
  {Rossolenko}, \citenamefont {Khabipov}, \citenamefont {Balashov},
  \citenamefont {Zorin}, \citenamefont {Dmitriev}, \citenamefont {Koshelets},\
  and\ \citenamefont {Ustinov}}]{feofanov2010implementation}%
  \BibitemOpen
  \bibfield  {author} {\bibinfo {author} {\bibfnamefont {A.}~\bibnamefont
  {Feofanov}}, \bibinfo {author} {\bibfnamefont {V.}~\bibnamefont {Oboznov}},
  \bibinfo {author} {\bibfnamefont {V.}~\bibnamefont {Bol'ginov}}, \bibinfo
  {author} {\bibfnamefont {J.}~\bibnamefont {Lisenfeld}}, \bibinfo {author}
  {\bibfnamefont {S.}~\bibnamefont {Poletto}}, \bibinfo {author} {\bibfnamefont
  {V.}~\bibnamefont {Ryazanov}}, \bibinfo {author} {\bibfnamefont
  {A.}~\bibnamefont {Rossolenko}}, \bibinfo {author} {\bibfnamefont
  {M.}~\bibnamefont {Khabipov}}, \bibinfo {author} {\bibfnamefont
  {D.}~\bibnamefont {Balashov}}, \bibinfo {author} {\bibfnamefont
  {A.}~\bibnamefont {Zorin}}, \bibinfo {author} {\bibfnamefont
  {P.}~\bibnamefont {Dmitriev}}, \bibinfo {author} {\bibfnamefont
  {V.}~\bibnamefont {Koshelets}},\ and\ \bibinfo {author} {\bibfnamefont
  {A.}~\bibnamefont {Ustinov}},\ }\bibfield  {title} {\bibinfo {title}
  {Implementation of superconductor/ferromagnet/ superconductor $\pi$-shifters
  in superconducting digital and quantum circuits},\ }\href
  {https://doi.org/10.1038/nphys1700} {\bibfield  {journal} {\bibinfo
  {journal} {Nat. Phys.}\ }\textbf {\bibinfo {volume} {6}},\ \bibinfo {pages}
  {593 EP } (\bibinfo {year} {2010})}\BibitemShut {NoStop}%
\bibitem [{\citenamefont {Hilgenkamp}(2008)}]{hilgenkamp2008pi}%
  \BibitemOpen
  \bibfield  {author} {\bibinfo {author} {\bibfnamefont {H.}~\bibnamefont
  {Hilgenkamp}},\ }\bibfield  {title} {\bibinfo {title} {Pi-phase shift
  josephson structures},\ }\href
  {http://stacks.iop.org/0953-2048/21/i=3/a=034011} {\bibfield  {journal}
  {\bibinfo  {journal} {Supercond. Sci. and Tech.}\ }\textbf {\bibinfo {volume}
  {21}},\ \bibinfo {pages} {034011} (\bibinfo {year} {2008})}\BibitemShut
  {NoStop}%
\bibitem [{\citenamefont {Andreanov}\ and\ \citenamefont
  {Fistul}(2018)}]{andreanov2017resonant}%
  \BibitemOpen
  \bibfield  {author} {\bibinfo {author} {\bibfnamefont {A.}~\bibnamefont
  {Andreanov}}\ and\ \bibinfo {author} {\bibfnamefont {M.~V.}\ \bibnamefont
  {Fistul}},\ }\bibfield  {title} {\bibinfo {title} {Resonant frequencies and
  spatial correlations in frustrated arrays of josephson type nonlinear
  oscillators},\ }\Eprint {https://arxiv.org/abs/1801.03842} {arXiv:1801.03842
  [cond-mat.mes-hall]}  (\bibinfo {year} {2018})\BibitemShut {NoStop}%
\bibitem [{\citenamefont {Ranadive}\ \emph {et~al.}(2022)\citenamefont
  {Ranadive}, \citenamefont {Esposito}, \citenamefont {Planat}, \citenamefont
  {Bonet}, \citenamefont {Naud}, \citenamefont {Buisson}, \citenamefont
  {Guichard},\ and\ \citenamefont {Roch}}]{ranadive2022kerr}%
  \BibitemOpen
  \bibfield  {author} {\bibinfo {author} {\bibfnamefont {A.}~\bibnamefont
  {Ranadive}}, \bibinfo {author} {\bibfnamefont {M.}~\bibnamefont {Esposito}},
  \bibinfo {author} {\bibfnamefont {L.}~\bibnamefont {Planat}}, \bibinfo
  {author} {\bibfnamefont {E.}~\bibnamefont {Bonet}}, \bibinfo {author}
  {\bibfnamefont {C.}~\bibnamefont {Naud}}, \bibinfo {author} {\bibfnamefont
  {O.}~\bibnamefont {Buisson}}, \bibinfo {author} {\bibfnamefont
  {W.}~\bibnamefont {Guichard}},\ and\ \bibinfo {author} {\bibfnamefont
  {N.}~\bibnamefont {Roch}},\ }\bibfield  {title} {\bibinfo {title} {Kerr
  reversal in josephson metamaterial and traveling wave parametric
  amplification},\ }\href@noop {} {\bibfield  {journal} {\bibinfo  {journal}
  {Nature communications}\ }\textbf {\bibinfo {volume} {13}},\ \bibinfo {pages}
  {1737} (\bibinfo {year} {2022})}\BibitemShut {NoStop}%
\bibitem [{\citenamefont {Planat}\ \emph {et~al.}(2019)\citenamefont {Planat},
  \citenamefont {Al-Tavil}, \citenamefont {Mart{\'\i}nez}, \citenamefont
  {Dassonneville}, \citenamefont {Foroughi}, \citenamefont {L{\'e}ger},
  \citenamefont {Bharadwaj}, \citenamefont {Delaforce}, \citenamefont
  {Milchakov}, \citenamefont {Naud} \emph {et~al.}}]{planat2019fabrication}%
  \BibitemOpen
  \bibfield  {author} {\bibinfo {author} {\bibfnamefont {L.}~\bibnamefont
  {Planat}}, \bibinfo {author} {\bibfnamefont {E.}~\bibnamefont {Al-Tavil}},
  \bibinfo {author} {\bibfnamefont {J.~P.}\ \bibnamefont {Mart{\'\i}nez}},
  \bibinfo {author} {\bibfnamefont {R.}~\bibnamefont {Dassonneville}}, \bibinfo
  {author} {\bibfnamefont {F.}~\bibnamefont {Foroughi}}, \bibinfo {author}
  {\bibfnamefont {S.}~\bibnamefont {L{\'e}ger}}, \bibinfo {author}
  {\bibfnamefont {K.}~\bibnamefont {Bharadwaj}}, \bibinfo {author}
  {\bibfnamefont {J.}~\bibnamefont {Delaforce}}, \bibinfo {author}
  {\bibfnamefont {V.}~\bibnamefont {Milchakov}}, \bibinfo {author}
  {\bibfnamefont {C.}~\bibnamefont {Naud}}, \emph {et~al.},\ }\bibfield
  {title} {\bibinfo {title} {Fabrication and characterization of aluminum squid
  transmission lines},\ }\href@noop {} {\bibfield  {journal} {\bibinfo
  {journal} {Physical Review Applied}\ }\textbf {\bibinfo {volume} {12}},\
  \bibinfo {pages} {064017} (\bibinfo {year} {2019})}\BibitemShut {NoStop}%
\bibitem [{\citenamefont {Kosterlitz}\ and\ \citenamefont
  {Thouless}(1973)}]{kosterlitz1973ordering}%
  \BibitemOpen
  \bibfield  {author} {\bibinfo {author} {\bibfnamefont {J.~M.}\ \bibnamefont
  {Kosterlitz}}\ and\ \bibinfo {author} {\bibfnamefont {D.~J.}\ \bibnamefont
  {Thouless}},\ }\bibfield  {title} {\bibinfo {title} {Ordering, metastability
  and phase transitions in two-dimensional systems},\ }\href@noop {} {\bibfield
   {journal} {\bibinfo  {journal} {Journal of Physics C: Solid State Physics}\
  }\textbf {\bibinfo {volume} {6}},\ \bibinfo {pages} {1181} (\bibinfo {year}
  {1973})}\BibitemShut {NoStop}%
\bibitem [{\citenamefont {Blinc}(1960)}]{blinc1960isotopic}%
  \BibitemOpen
  \bibfield  {author} {\bibinfo {author} {\bibfnamefont {R.}~\bibnamefont
  {Blinc}},\ }\bibfield  {title} {\bibinfo {title} {On the isotopic effects in
  the ferroelectric behaviour of crystals with short hydrogen bonds},\ }\href
  {https://doi.org/https://doi.org/10.1016/0022-3697(60)90003-2} {\bibfield
  {journal} {\bibinfo  {journal} {Journal of Physics and Chemistry of Solids}\
  }\textbf {\bibinfo {volume} {13}},\ \bibinfo {pages} {204} (\bibinfo {year}
  {1960})}\BibitemShut {NoStop}%
\bibitem [{\citenamefont {Trotter}(1959)}]{trotter1959product}%
  \BibitemOpen
  \bibfield  {author} {\bibinfo {author} {\bibfnamefont {H.~F.}\ \bibnamefont
  {Trotter}},\ }\bibfield  {title} {\bibinfo {title} {On the product of
  semi-groups of operators},\ }\href {http://www.jstor.org/stable/2033649}
  {\bibfield  {journal} {\bibinfo  {journal} {Proceedings of the American
  Mathematical Society}\ }\textbf {\bibinfo {volume} {10}},\ \bibinfo {pages}
  {545} (\bibinfo {year} {1959})}\BibitemShut {NoStop}%
\bibitem [{\citenamefont {Baxter}(2016)}]{baxter2016exactly}%
  \BibitemOpen
  \bibfield  {author} {\bibinfo {author} {\bibfnamefont {R.~J.}\ \bibnamefont
  {Baxter}},\ }\href@noop {} {\emph {\bibinfo {title} {Exactly solved models in
  statistical mechanics}}}\ (\bibinfo  {publisher} {Elsevier},\ \bibinfo {year}
  {2016})\BibitemShut {NoStop}%
\bibitem [{\citenamefont {Strecka}\ and\ \citenamefont
  {Jascur}(2015)}]{strecka2015brief}%
  \BibitemOpen
  \bibfield  {author} {\bibinfo {author} {\bibfnamefont {J.}~\bibnamefont
  {Strecka}}\ and\ \bibinfo {author} {\bibfnamefont {M.}~\bibnamefont
  {Jascur}},\ }\href {https://arxiv.org/abs/1511.03031} {\bibinfo {title} {A
  brief account of the ising and ising-like models: Mean-field, effective-field
  and exact results}} (\bibinfo {year} {2015}),\ \Eprint
  {https://arxiv.org/abs/1511.03031} {arXiv:1511.03031 [cond-mat.stat-mech]}
  \BibitemShut {NoStop}%
\bibitem [{\citenamefont {Suzuki}(1985)}]{suzuki1985transfer}%
  \BibitemOpen
  \bibfield  {author} {\bibinfo {author} {\bibfnamefont {M.}~\bibnamefont
  {Suzuki}},\ }\bibfield  {title} {\bibinfo {title} {Transfer-matrix method and
  monte carlo simulation in quantum spin systems},\ }\href
  {https://doi.org/10.1103/PhysRevB.31.2957} {\bibfield  {journal} {\bibinfo
  {journal} {Phys. Rev. B}\ }\textbf {\bibinfo {volume} {31}},\ \bibinfo
  {pages} {2957} (\bibinfo {year} {1985})}\BibitemShut {NoStop}%
\end{thebibliography}%

\end{document}